\documentclass[12pt,english]{article}

\usepackage[a4paper,lmargin=70pt,rmargin=70pt]{geometry}
\usepackage{setspace}
\usepackage{amsthm}
\usepackage{amsmath}
\usepackage{amssymb}
\usepackage{tensor}
\usepackage{babel}
\usepackage{graphicx}

\usepackage{setspace}
\usepackage{subfig}
\usepackage{dsfont}
\usepackage{cite}

\usepackage[colorlinks=true
,urlcolor=blue
,anchorcolor=blue
,citecolor=blue
,filecolor=blue
,linkcolor=blue
,menucolor=blue
,linktocpage=true
]{hyperref}

\newcommand{\dho}{\partial}

\newcommand{\Hence}{\quad\Longrightarrow\quad}
\newcommand{\ed}{\,.}
\newcommand{\ec}{\,,}
\newcommand{\ecq}{\ec\quad}
\newcommand{\ind}[1]{\indices{#1}}
\newcommand{\bR}{\ensuremath{\mathbb{R}}}
\newcommand{\bZ}{\ensuremath{\mathbb{Z}}}
\newcommand{\cM}{\ensuremath{\mathcal{M}}}
\newcommand{\cO}{\ensuremath{\mathcal{O}}}
\newcommand{\ttheta}{\tilde{\theta}}
\newcommand{\tf}{\tilde{f}}
\newcommand{\tz}{\tilde{z}}
\newcommand{\txi}{\tilde{\xi}}
\newcommand{\gint}{g_{(6)}}
\newcommand{\Gten}{G^{(10)}}
\newcommand{\Gfour}{G^{(4)}}
\newcommand{\GfourE}{G^{(4)E}}
\newcommand{\bg}{\mathrm{bg}}
\newcommand{\kten}{\kappa_{10}}
\newcommand{\tmu}{\tilde{\mu}}
\newcommand{\Rten}{R^{(10)}}
\newcommand{\Rfour}{R^{(4)}}
\newcommand{\RfourE}{R^{(4)E}}
\newcommand{\Mp}{M_{\mathrm{pl}}}
\newcommand{\kin}{\mathrm{kinetic}}
\newcommand{\ap}{\alpha'}
\newcommand{\Dp}{\mathrm{D}p}
\newcommand{\Op}{\mathrm{O}p}
\newcommand{\Osix}{\mathrm{O}6}
\newcommand{\Dsix}{\mathrm{D}6}
\newcommand{\Dfour}{\mathrm{D}4}
\newcommand{\KK}{\mathrm{KK}}
\newcommand{\KKfive}{\mathrm{KK}5}
\newcommand{\NSfive}{\mathrm{NS}5}
\newcommand{\Span}{\mathrm{Span}}
\newcommand{\IIA}{\mathrm{IIA}}
\newcommand{\wv}{\mathrm{w.v.}}
\newcommand{\flux}{\mathrm{flux}}
\newcommand{\tot}{\mathrm{total}}

\date{}

\begin{document}

\title{
\vspace{-20mm}
\hfill{\small \tt WIS/13/13-OCT-DPPA}\vskip 5pt
Brane Inflation and Moduli Stabilization \\ on Twisted Tori}
\author{Guy Gur-Ari\\\\
{\it Department of Particle Physics and Astrophysics}\\
{\it Weizmann Institute of Science, Rehovot 76100, Israel}\\
{\small{\tt e-mail:~Guy.GurAri@weizmann.ac.il}}
}
\maketitle

\begin{abstract}
  We consider supergravity compactifications on 6-dimensional twisted tori, which are 5-torus fibrations of the circle.
  The motion of branes on such manifolds can lead to power-law potentials at low energy, that may be useful for inflation.
  We classify the possible low energy potentials one can obtain by wrapping branes on different cycles of the fibre.
  Turning to the problem of moduli stabilization in such models, we prove a no-go result for solutions with parametrically small cosmological constant, under certain assumptions for the orientifolds and D-branes.
  We also consider the role of discrete Wilson lines in moduli stabilization on general closed manifolds, and show that gauge invariance restricts their contributions to the effective potential.
  We derive the allowed discrete Wilson lines in massive Type IIA supergravity on twisted tori.
  We conclude with a detailed example, computing the effective potentials in a class of models involving a twisted torus and an orientifold 6-plane.
\end{abstract}

\tableofcontents{}

\setlength{\parindent}{0pt}
\setlength{\parskip}{1ex}

\newpage

\section{Introduction}

In string theory, it is notoriously difficult to construct solutions with positive cosmological constant.
Such solutions may be useful for describing the epoch of inflation, as well as the observed acceleration of the present-day universe.
In this note we consider supergravity compactifications on twisted tori, and  discuss obstacles to obtaining de Sitter vacua in such compactifications. 
Twisted tori have been proposed as useful backgrounds for brane inflation \cite{Silverstein:2008sg}, and we explore the possible inflaton potentials one can obtain on them.

The conceptually simplest models of inflation involve a single inflaton field $\varphi$ with a shallow power-law potential $V(\varphi) \sim \varphi^s$ (see \cite{Baumann:2009ds} for a review of inflation).
The inflaton is given a large initial expectation value and it rolls slowly down the potential, leading to inflation.
When the inflaton approaches the bottom of the potential, inflation ends and reheating occurs.
In these models, one can show that the initial expectation value must be larger than the Planck scale, in order to generate enough $e$-foldings to explain cosmological observations.
For this reason these models are called `large-field models'.
They are interesting in part due to the Lyth bound \cite{Lyth:1996im}, which states that a significant tensor-to-scalar ratio requires super-Planckian inflaton values (see \cite{Hotchkiss:2011gz} for a suggestion on evading the Lyth bound).

Unfortunately, a model that involves super-Planckian field values cannot be completely described within the confines of quantum field theory.
If we assume that field theory breaks down at or below the Planck scale, then the best we can hope for is to write down an effective field theory with a cutoff at the Planck scale. 
Such a theory must include an infinite number of Planck-suppressed operators of the form $\varphi^{k+4} / \Mp^k$, all of which become important when taking $\langle \varphi \rangle > \Mp$.
Within this framework, simple power-law models therefore require the fine-tuning of an infinite number of couplings.
One may address this problem by invoking an approximate shift symmetry of the inflaton, a symmetry that is weakly broken by the inflaton's potential.
Still, one must verify the presence of such a symmetry in a UV completion of gravity.

Notice that this problem is far worse than the more common `eta problem' of slow-roll inflation.
There, Planck-suppressed contributions to the inflaton mass of the form $\cO_4 \varphi^2/ \Mp^2$ can lead to a violation of the slow-roll conditions, which are necessary for maintaining the exponential growth of the universe during inflation.
In this context, a fine-tuning of a more manageable number of parameters may be required to produce inflation.

When working within a UV-complete framework such as string theory, there is at least a chance of achieving a complete realization of power-law inflation, in which all corrections to the inflaton potential are under control.
This approach comes with its own set of problems.
In a string theory realization, the inflaton is usually the low-energy description of a string theory object that lives on the internal manifold.
The fact that this manifold is compact often makes it difficult to extend the inflaton field range beyond the Planck scale.
For example, in brane-anti-brane models \cite{Kachru:2003sx,Burgess:2001fx,Dvali:2001fw}, the inflaton is the distance between a brane and an anti-brane that live on the compact manifold.
The field range is then directly limited by the size of the manifold.

In \cite{Silverstein:2008sg}, Silverstein and Westphal suggested a way of using monodromy to extend the field range and solve the eta problem (see also \cite{McAllister:2008hb}).
To illustrate the idea, let us take the internal manifold to be a twisted torus, namely a fibration of a 2-torus over a circle (the remaining orthogonal directions on the manifold will not be important for us).
Let $z$ denote the coordinate along the circle, and let $x,y$ denote the canonical coordinates of the 2-torus, with periodicity 1.
The twisted torus is defined by the identification
\begin{align}
  (z,x,y) \cong (z-1,x+y,y) \ed
\end{align}
In other words, when making a rotation around the circle, the fibre is transformed by the monodromy $T=\begin{pmatrix}
    1 & 1 \\ 0 & 1
\end{pmatrix}$ of $SL(2,\bZ)$.

Now, consider a space-filling D4-brane (the `inflaton brane') that is localized on the $z$-circle, and that wraps the $y$-cycle of the fibre.
When the brane makes a rotation around the $z$ direction,
the cycle it wraps undergoes a $T$ transformation, causing it to wind further around the $x$-cycle.
Due to the monodromy, the field $\varphi$ that represents the position of the brane on the circle is no longer compact. One can show that the monodromy, which causes the brane to stretch, leads to an effective potential $V(\varphi) \sim \varphi^{2/3}$ for this field.

In \cite{Silverstein:2008sg} the inflaton brane was treated as a probe, and the moduli\footnote{By `moduli' we shall mean both pseudo-moduli, whose effective potential is not flat, and actual moduli.} of the twisted torus background were stabilized as in \cite{Silverstein:2007ac}, by introducing orientifold planes, fluxes, discrete Wilson lines, and Kaluza-Klein monopoles.
The background was shown to have low curvature and small coupling, and therefore both stringy and quantum corrections were under control.
This was achieved without invoking supersymmetry.

This construction has another interesting aspect, in that the background of \cite{Silverstein:2007ac} is by itself a proposed supergravity solution with positive cosmological constant; for additional results regarding inflation and de Sitter vacua in Type IIA supergravity, see 
\cite{Hertzberg:2007wc,Caviezel:2008ik,Caviezel:2008tf,Flauger:2008ad,Danielsson:2009ff,deCarlos:2009fq,deCarlos:2009qm,Wrase:2010ew,Danielsson:2010bc,Andriot:2010ju,Danielsson:2011au}.
Such solutions are much more difficult to obtain in string theory than those with negative cosmological constant.
It is easy to see part of the difficulty already in the supergravity limit.
As we review below, in the smeared approximation the effective 4D potential of the moduli takes the form
\begin{align}
  U = a g^2 + b g^3 + c g^4 \ec
  \label{Uintro}
\end{align}
where $g \sim e^\phi$, $\phi$ is the dilaton, and $a$, $b$, $c$ are functions of the remaining moduli, which we assume to be stabilized.
When $c$ is positive, any minimum with negative $a$ is an anti-de Sitter vacuum. The Freund-Rubin solution \cite{Freund:1980xh} is of this type: it plays the $g^2$ contribution from the sphere's positive curvature off the $g^4$ contribution of a Ramond-Ramond flux through the sphere.
Constructing a de Sitter solution is more complicated, because it requires a delicate balance between $a$, $b$, and $c$ to form a metastable solution with positive $U$.

In this note we generalize the work of Silverstein and Westphal to internal manifolds that are 5-torus fibrations of $S^1$, with general $SL(5,\bZ)$ monodromies.
After introducing twisted tori in section \ref{TTG}, we discuss brane inflation on these models in section \ref{BI}, ignoring for the moment the issue of moduli stabilization.
We classify the possible low-energy potentials $V(\varphi)$ that one can obtain by wrapping space-filling D-branes around fibre cycles, where $\varphi$ is the brane position along the (cover of) $S^1$.
We find that the possibilities for asymptotic behavior at large $\varphi$ are
\begin{align}
  V(\varphi) \sim \varphi^{2/3}, \, \varphi, \, \varphi^{6/5}, 
  \, \varphi^{4/3}, \, 
  \varphi^{10/7}, \, \varphi^{3/2}, \, \varphi^2 \ed
  \label{potentials}
\end{align}
Most of these potentials are obtained by generalizing the $T \in SL(2,\bZ)$ transformation to larger Jordan blocks in $SL(5,\bZ)$ transformations.

We then address the problem of moduli stabilization in twisted torus models, within the context of Type IIA supergravity, which we review in section \ref{IIA1}.
Twisted torus compactifications have been previously considered in 
\cite{Scherk:1979zr,Bergshoeff:1996ui,Cowdall:1996tw,Kaloper:1999yr,Cvetic:2003jy,Gurrieri:2002wz,Gurrieri:2002iw,Schulz:2004ub,Schulz:2004tt,
Derendinger:2004jn,
Derendinger:2005ph,
Dall'Agata:2005ff,Villadoro:2005cu,Camara:2005dc,Hull:2005hk,Fre':2006ut,ReidEdwards:2009nu}.

In section \ref{nogo} we discuss vacua with parametrically small cosmological constant in twisted torus compactifications.
As was shown in \cite{Silverstein:2007ac,Haque:2008jz}, stabilizing the potential \eqref{Uintro} when the cosmological constant is parametrically small is equivalent to minimizing the determinant $\Delta = 4ac/b^2$.
We show that the determinants in twisted torus compactifications of massive Type IIA have a runaway direction when all the orientifolds and D-branes wrap the base circle. This rules out vacua with parametrically small cosmological constants in such models.
The instability can also appear when the compactification manifold is a product of two twisted tori; in particular, the model that was suggested in \cite{Silverstein:2007ac} suffers from this instability.

We will work in the approximation where all localized sources are smeared across the internal manifold, and for a given solution it is important to check whether this approximation is valid.
In \cite{Douglas:2010rt} it was shown that the smeared approximation can break down when the internal manifold has negative curvature, and this includes twisted tori.
In some cases, however, it is enough to turn on weak warping to evade this no-go result, and then the smeared approximation is valid \cite{Dong:2010pm}. See also \cite{Blaback:2010sj} for a discussion of this issue in the context of T-duality.

Twisted tori generally include torsion cycles, which are cycles of finite rank in the homology.
Wilson lines around such cycles are discretized.
In section \ref{DWL} we discuss the role of discrete Wilson lines in moduli stabilization of supergravity models.
The supergravity field strengths include Chern-Simons-like terms that are affected by Wilson-line degrees of freedom. 
For example, in massive Type IIA supergravity the RR field strengths are given by
\begin{align}
  \tilde{F}_2 &= dC_1 + m_0 B_2 \ecq
  \tilde{F}_4 = dC_3 - C_1 \wedge H_3
  - \frac{m_0}{2} B_2 \wedge B_2 \ed
\end{align}
The potential energies $|\tilde{F}|^2$ are therefore sensitive, in general, to values of $C_1$ Wilson lines and $B_2$ `Wilson surfaces' (integrals of $B_2$ over 2-cycles).
One may then expect that discrete Wilson lines, which are Wilson lines placed on torsion cycles, can be used for moduli stabilization in a way similar to fluxes.

We show that the effect of discrete Wilson lines on the field strengths is restricted by gauge invariance.
Discrete $B_2$ Wilson surfaces cannot contribute to $\tilde{F}_2$, and there can be no terms in the field strengths that involve both discrete Wilson lines and background field fluctuations.
The only allowed contributions are discrete, and may arise from the terms $C_1 \wedge H_3$ and $m_0 B_2 \wedge B_2$. 
They are allowed only when these terms are globally defined forms.
The contributions of discrete Wilson lines are therefore similar to those of ordinary $\tilde{F}_4$ fluxes.
An analysis of the gauge transformations suggests that these contributions may shift the $\tilde{F}_4$ fluxes by non-integer values.
In particular, in the class of manifolds we are interested in, the analysis suggests that a discrete $C_1$ Wilson line of rank $N$ can shift the flux by (see section \ref{IIA1} for our conventions)
\begin{align}
  \sqrt{2} \int \tilde{F}_4 \in \frac{(2\pi)^3}{N} \bZ \ed
\end{align}
And, when $B_2$ is a sum of two discrete Wilson surfaces of ranks $N$, $M$, the contribution to the flux is in
\begin{align}
  \sqrt{2} \int \tilde{F}_4
  \in \frac{(2\pi)^3}{\gcd(N,M)} \bZ \ed
\end{align}
These results follow from an analysis of the supergravity gauge transformations on twisted tori.
$p$-form gauge theories include higher-order gauge transformations; these are gauge transformations of the gauge parameters.
A complete analysis, which is beyond the scope of this work, must include also these transformations; it may reveal additional obstructions, and may modify the results for the non-integer shifts.
Nevertheless, these results do affect some of the discrete Wilson line contributions that were used in \cite{Silverstein:2007ac} for stabilization.
In particular, the contributions arising from the $|\tilde{F}_2|^2$ term in the potential are ruled out.

Despite the above obstacles to the construction of de Sitter vacua in twisted torus models, we have searched for such vacua in a class of models involving different monodromies and including orientifold planes, fluxes, and Kaluza-Klein monopoles, in the context of Type IIA supergravity.
We have not yet been able to find such solutions.
In section \ref{expl} we present a detailed example of one class of potentials involving orientifold 6-planes.

Additional details are included in several appendices. Appendix \ref{TTGapp} complements section \ref{TTG} and includes details on the geometry of twisted tori. Appendix \ref{IIA} is a review of massive Type IIA supergravity, including its democratic formulation. Appendix \ref{DWLtori} includes various derivations and examples of discrete Wilson lines on twisted tori, complementing section \ref{DWL}.

\section{Twisted Torus Geometry}
\label{TTG}

A twisted $(n+1)$-torus is a manifold $\cM$, with coordinates $z, \xi^a$, $a=1,\dots,n$, and with the identifications
\begin{align}
  \xi^a \cong \xi^a + 1 \ecq
  (z,\xi^a) \cong (z-1,M\ind{^a_b} \xi^b) \ecq
  M \in SL(n,\bZ) \ed
  \label{ident}
\end{align}
One can think of this manifold as a fibration of an $n$-torus over a circle, where $z$ is the coordinate along the circle.
The first identification creates the $n$-torus fibre at each $z$ value. The second identification states that when we make a roundtrip around the circle, the $n$-torus is glued back to itself under a mapping class group transformation $M$; we call $M$ the monodromy.
See \cite{Hull:2005hk} for a discussion of the geometry of these manifolds.

Let us assume there is a real matrix $X$ such that $M = e^X$. This will be true for all the cases we will be interested in.\footnote{See \cite{Culver:1966aa} for a precise criterion.}
With this assumption we can write down a set of independent real 1-forms $\{ \theta^A \}$,\footnote{We use upper-case roman letters to denote the full set of coordinates on $\cM$, and lower-case roman letters for the fibre coordinates. Greek letters denote spacetime indices.}
\begin{align}
  \theta^z = dz \ecq
  \theta^a = \gamma(z)\ind{^a_b} d\xi^b \ecq
  \gamma(z) = e^{zX} \ed
  \label{frame}
\end{align}
These forms are globally defined and nowhere-vanishing, so they form a global frame on $\cM$.
They obey a Cartan structure equation
\begin{align}
  d\theta^A + \frac{1}{2} f\ind{_B_C^A} \, \theta^B \wedge \theta^C = 0 \ec
  \label{cartan}
\end{align}
where the non-vanishing components of $f$ are $f\ind{_b_z^a} = - f\ind{_z_b^a} = X_{ab}$.
This gives $\cM$ a local group manifold structure, and globally $\cM$ is a coset space.

The global frame defines a metric on $\cM$. For purposes of moduli stabilization we will allow diagonal components on this metric to vary, but will keep off-diagonal components at zero.
We therefore take the metric to be
\begin{align}
  \gint
  &= L_z^2 (\theta^z)^2 + \sum_{a=1}^n L_a^2 (\theta^a)^2 
  = L_z^2 dz^2 + d\vec{\xi}^T \gamma(z)^T L^2 \gamma(z) d\vec{\xi} \ec
  \label{g6}
\end{align}
where $L_A$ measure cycle lengths, and $L = \mathrm{diag}(L_1,\dots,L_n)$.
The volume is
\begin{align}
  V &= \int_0^1 dz d^n\xi \sqrt{\gint} = L_z L_1 \cdots L_n 
  \label{V} \ed
\end{align}
For reasons that will become clear in the next section, in what follows we will assume that $X$ is strictly upper-triangular, with zeros on the diagonal. In this case the curvature, which is constant and negative, is
\begin{align}
  R &= - \frac{1}{2 L_z^2} \sum_{a<b} \left( \frac{L_a}{L_b} X_{ab} \right)^2
  \label{R} \ed
\end{align}

These are the basic properties we will need in order to proceed.
Other geometric properties are derived in appendix \ref{TTGapp}.

\section{Brane Inflation}
\label{BI}

In this section we consider a space-filling D-brane in a background of $\bR^{1,3} \times \cM$, where $\cM$ is a twisted 6-torus.
The brane is localized in the $z$ direction, and wraps a $p$-cycle on the internal manifold.

Silverstein and Westphal \cite{Silverstein:2008sg} considered a similar setup, based on a twisted 3-torus. 
They showed that the position of the brane along the $z$
direction can act, at low energies, as a scalar field with a shallow power-law
potential.
The brane gets twisted as it moves around the $z$ direction and this extends the inflaton target space, even though the $z$ direction itself is compact.
As long as all the moduli are stabilized, and the backreaction of the inflaton brane is small, this gives an implementation of large-field inflation in which all corrections are under control.\footnote{In some cases the backreaction from the inflaton's coupling to heavy fields can, in fact, help flatten the inflaton's potential \cite{Dong:2010in}.}

In this section we generalize the construction of \cite{Silverstein:2008sg} and compute the possible power-law potentials one can get when using different monodromies and D-branes. We will be interested only in the scaling properties of the potential at large field values, so for simplicity let us set $L_z=L_i=1$ until the end of the section.

To begin with, consider a D4-brane that fills spacetime, is initially localized at $z=0$, and that wraps a 1-cycle $\sigma = \sigma_a d\xi^a$ of the 5-torus fibre.
Let $x^\mu$ denote the spacetime coordinates, and let $\zeta\in[0,1)$ denote the coordinate along the $\sigma$ cycle. The brane's embedding on the $T^5$ is given by $\xi^a(\zeta) = \sigma_a \zeta$, $\sigma_a \in \bZ$.

Suppose we push the brane uniformly in the $z$ direction, along the base circle.
Its motion will not be periodic in general, because with each rotation the cycle that it wraps will be altered by the monodromy.
It is convenient to lift the brane to $\bR^{1,3} \times \bR \times T^5$ with the metric \eqref{g6}, where $z$ is now the coordinate along the $\bR$ factor.
Denote the brane's position along the lifted $z$ direction by $\tilde{\varphi}(x)$; this will be the inflaton field, though with this definition it will have a non-canonical kinetic term.
As the brane moves along the $z$ direction the metric \eqref{g6} will cause it to stretch, leading to an effective potential for the low-energy inflaton.

The induced metric on the brane is
\begin{align}
  g_{\rm ind.} &= 
  \left( \eta_{\mu\nu} + \dho_\mu \tilde{\varphi} \dho_\nu \tilde{\varphi} \right) 
  dx^\mu dx^\nu + 
  \left( \sigma^T e^{\tilde{\varphi} X^T} e^{\tilde{\varphi} X} \sigma \right) d\zeta^2 \ec
\end{align}
where $\eta$ is the Minkowski metric, and $\sigma$ is a vector with components $\sigma_a$.
To obtain the effective potential for the field $\tilde{\varphi}(x)$, we write down the DBI action for the brane and make a derivative expansion in $\dho_\mu\tilde{\varphi}$. Here we are assuming the brane moves slowly and with small gradients, which is consistent with the slow-roll conditions of large-field inflation.
\begin{align}
  S_{\Dfour} &= -\tau_4 \int \! d^4x \, d\zeta \,
  \sqrt{-\det g_{\rm ind.}}
  = -\tau_4 \int \! d^4x 
  \left[ 
  \tilde{V}(\tilde{\varphi}) + 
  \frac{1}{2} \tilde{V}(\tilde{\varphi}) \dho_\mu\tilde{\varphi} \dho^\mu\tilde{\varphi} 
  + O(\dho^4) 
  \right] 
  \ec
  \label{D4action}
\end{align}
where $\tilde{V}(\tilde{\varphi}) = \left|e^{\tilde{\varphi} X} \sigma\right|$.

Next, make a field redefinition $\varphi = \varphi(\tilde{\varphi})$ and demand that $\varphi$ have a canonical kinetic term, so that the action becomes
\begin{align}
   -\tau_4 \int \! d^4x \,
  \left[ \frac{1}{2} \dho_\mu\varphi \dho^\mu\varphi + V(\varphi) + O(\dho^4)
  \right] \ed
\end{align}
It is easy to see that the field $\varphi$ and the canonical potential $V$ are given by
\begin{align}
  \varphi(\tilde{\varphi}) = \int^{\tilde{\varphi}}
  \! d\tilde{\varphi}' \, \tilde{V}^{1/2}(\tilde{\varphi}') \ecq
  V(\varphi) = \tilde{V}\left( \tilde{\varphi}(\varphi) \right) \ed
  \label{canV}
\end{align}

Let us see which kinds of non-trivial inflaton potentials $V(\varphi)$ one can get for different choices of monodromy $e^X$ and brane cycle $\sigma$.
Here we are only interested in the scaling behavior of the potential at large $\varphi$, and therefore at large $\tilde{\varphi}$.
If $\sigma$ is an eigenvector of $e^X$ with eigenvalue $\lambda \ne 1$, then the non-canonical potential will scale as $\tilde{V}(\tilde{\varphi}) \sim |\lambda^{\tilde{\varphi}} \sigma|$, the canonical inflaton will be $\varphi \sim |\lambda|^{\tilde{\varphi}/2}$, and the resulting canonical potential will go as $V(\varphi) \sim \varphi^2$ at large $\varphi$.

To obtain additional potentials, suppose the monodromy $e^X$ is upper-triangular with ones on the diagonal, and general entries above it. Raising it to the power $\tilde{\varphi}$, it will generically scale as
\begin{align}
  e^{\tilde{\varphi} X} \sim
  \begin{pmatrix}
    1 & \tilde{\varphi} & \tilde{\varphi}^2 & \tilde{\varphi}^3 & \tilde{\varphi}^4 \\
      & 1            & \tilde{\varphi}   & \tilde{\varphi}^2 & \tilde{\varphi}^3 \\
      &              & 1              & \tilde{\varphi}   & \tilde{\varphi}^2 \\
      &              &                & 1              & \tilde{\varphi}   \\
      &              &                &                & 1
  \end{pmatrix} \ed
\end{align}
Different choices of the cycle $\sigma$ will then lead to $\tilde{V}$ scaling as $\tilde{\varphi}^s$ for $s=1,2,3,4$. 
The canonical potentials \eqref{canV} in these cases go as
\begin{align}
  V(\varphi) \sim \varphi^{\frac{2s}{s+2}}
  \label{Vvarphi}
\end{align}
at large $\varphi$. Specifically, we can get potentials that go as $\varphi^{2/3},\varphi,\varphi^{6/5},\varphi^{4/3}$.

More generally, we can let $e^X$ be block-diagonal, and have one of the blocks take this upper-triangular form. The potentials we can get will then be limited by the size of the block.
For example, the potential $\varphi^{2/3}$ (which was already obtained in \cite{Silverstein:2008sg}) corresponds to a $2 \times 2$ upper-triangular block in the monodromy matrix.

Two more potentials can be obtained by using a D5-brane that wraps a 2-cycle on the torus. If this 2-cycle factorizes as $(\sigma^{(1)}_a d\xi^a) \wedge (\sigma^{(2)}_b d\xi^b)$, we can denote it as $\sigma_2 = ( \sigma^{(1)} \,\, \sigma^{(2)} )$ where $\sigma^{(i)}$ are column vectors.
In this case the brane action takes the form \eqref{D4action} but now with $\tilde{V}(\tilde{\varphi}) = 
\sqrt{\det (\sigma_2^T e^{\tilde{\varphi} X^T} e^{\tilde{\varphi} X} \sigma_2)}$. By taking the monodromy $e^X$ to be a single upper-triangular block, we can obtain additional $\tilde{V} \sim \tilde{\varphi}^s$ potentials with powers
\begin{align}
  s=5 \, : \quad
  \sigma_2^T = \left( \begin{smallmatrix}
    0 & 0 & 1 & 0 & 0 \\
    0 & 0 & 0 & 0 & 1
  \end{smallmatrix} \right) \ec \\
  s=6 \, : \quad
  \sigma_2^T = \left( \begin{smallmatrix}
    0 & 0 & 0 & 1 & 0 \\
    0 & 0 & 0 & 0 & 1
  \end{smallmatrix} \right) \ed
\end{align}

One can show that there are no additional power-law potentials that can be obtained using D4- or D5-branes, for any monodromy that can be written as $e^X$ with real $X$ \cite{Gurari:2010aa}. We will not go into the details here, but only sketch the argument. One starts by considering the Jordan form of a general monodromy, and expanding the brane's 1- or 2-cycle in its Jordan basis. Power-law behavior of $\tilde{V}(\tilde{\varphi})$ corresponds to having non-trivial Jordan blocks with eigenvalue 1; exponential behavior (which is equivalent to a quadratic canonical potential) can come from eigenvalues whose absolute value is not 1.
A direct calculation then shows that power-law behavior of $\tilde{V}$ is always polynomial with maximal degree 6. We already obtained all such exponents explicitly.

D6- and D7-branes do not yield any new potentials. To see this, T-dualize along the $T^5$ directions. Because the background $B$ field vanishes, and the metric has no off-diagonal modes relating the $T^5$ and the other directions, the result is another $T^5$ with an inverse metric and with no background $B$ field.\footnote{The T-duality transformation $\cO \in O(5,5)$ acts on the sum of the vector and 1-form spaces of the $T^5$, or equivalently on momenta and winding numbers. Since the duality exchanges them in pairs, the relevant transformation is $\cO = \begin{pmatrix}
  0 & \mathds{1} \\ \mathds{1} & 0
\end{pmatrix}$. It acts on the generalized metric $H = \begin{pmatrix}
  g - B g^{-1} B & B g^{-1} \\ -g^{-1} B & g^{-1}
\end{pmatrix}$ as $H \mapsto \cO^T H \cO$ \cite{Grana:2008yw}, and we get $g \mapsto g^{-1}$.
}
Inverting the $T^5$ part of the metric $\eqref{g6}$ is equivalent to changing the monodromy by $X \mapsto -X^T$.
As for the branes, this T-duality maps D6- and D7-branes to D5- and D4-branes respectively. These T-dual cases were all analyzed above.

Let us summarize what we found so far.
The inflaton potentials with asymptotic power-law behavior that one can construct by placing D-branes on twisted tori, are given by $V(\varphi) \sim \varphi^{2s/(s+2)}$ where $s=1,2,\dots,6$.
All except for the quadratic potential can be constructed by taking the monodromy to be upper-triangular, with ones on the diagonal. 

In what follows, we will consider string theory on twisted tori with such monodromies, in the supergravity limit. The idea is to first stabilize all moduli in these backgrounds and obtain a solution with a small cosmological constant. If we manage to do this, we can then try to achieve inflation by introducing the inflaton brane in the probe approximation.

\section{Massive Type IIA Supergravity}
\label{IIA1}

In this section we summarize basic properties of massive Type IIA supergravity \cite{Romans:1985tz}, following \cite{DeWolfe:2005uu}. For a more detailed review, see appendix \ref{IIA}.

The theory includes a metric $\Gten$, a dilaton $\phi$, and $p$-form gauge fields $B_2$, $C_1$, and $C_3$. 
We consider only the bosonic part of the action, which can be written as
\begin{align}
  S = S_{\kin} + S_{\mathrm{CS}} \ed
\end{align}
In string frame, the kinetic piece is
\begin{align}
  S_{\kin} &= \frac{1}{2\kten^2}
  \int d^{10}x \sqrt{-\Gten} \left[
  e^{-2\phi} \left( \Rten + 4 (\dho_\mu \phi)^2 
  - \frac{1}{2} |H_3|^2 \right)
  - |\tilde{F}_2|^2
  - |\tilde{F}_4|^2
  - m_0^2
  \right] \ec
  \label{Skin}
\end{align}
where the gauge-invariant field strengths are
\begin{align}
  H_3 &= dB_2 \ec \label{IIAH3} \\
  \tilde{F}_2 &= dC_1 + m_0 B_2 \ec \label{IIAF2} \\
  \tilde{F}_4 &= dC_3 - C_1 \wedge H_3
  - \frac{m_0}{2} B_2 \wedge B_2 \label{IIAF4} \ec
\end{align}
and
\begin{gather}
  2\kten^2 = (2\pi)^7 \alpha'^4 \ecq
  |F_p|^2 = \frac{1}{p!} F_{\mu_1 \dots \mu_p} F^{\mu_1 \dots \mu_p} \ed
\end{gather}
The fields $B_2,C_1,C_3$ include both fluctuations and backgrounds that give rise to discrete fluxes.
The Chern-Simons piece $S_{\mathrm{CS}}$ is written down in appendix \ref{IIA}.
It will not be important for us, because we will focus on the effective potential of the dilaton and the diagonal metric modes.

Let us take the 10D manifold to be a product $\bR^{3,1} \times \cM$. The metric in string frame is written as $\Gten = \Gfour + \gint$, with $\Gfour$ the spacetime metric and $\gint$ the metric on $\cM$.
We assume there is no warping, so the 10D curvature is given by $\Rten = \Rfour + R$, where $\Rfour$ is the spacetime curvature and $R$ is the curvature of the internal manifold. 
We will further assume that the dilaton is constant on the internal manifold.

In the presence of D-branes, the action receives the additional contributions
\begin{align}
  S_{\Dp} &= - \mu_p \int d^{p+1}\zeta e^{-\phi} \sqrt{-g_{(p)}}
  + \sqrt{2} \mu_p \int C_{p+1} + \cdots \ec 
  \label{SDp}
\end{align}
where the integrals are over the brane's worldvolume.
For orientifold planes the action is the same except that we replace $\mu_p$ by the orientifold charge $\tmu_p$.
The additional terms in \eqref{SDp}, which will not be important for us, involve both the bulk and brane gauge fields and are necessary for maintaining gauge invariance and invariance under dualities \cite{Green:1996bh}. 
The D-brane/O-plane charges (and tensions) are
\begin{align}
  \mu_p &= (2\pi)^{-p} (\alpha')^{-\frac{p+1}{2}} \ec \\
  \tmu_p &= - 2^{p-5} \mu_p \ed
\end{align}

We can turn on fluxes for $H_3$, $\tilde{F}_2$, and $\tilde{F}_4$, subject to the quantization conditions
\begin{align}
  \int H_3 &= 2 \kten^2 \mu_5 h_3
  = (2\pi)^2 \alpha' h_3 \ecq
  h_3 \in \bZ \ec
  \notag \\
  \sqrt{2} \int \tilde{F}_p &= 2 \kten^2 \mu_{8-p} f_p 
  = (2\pi)^{p-1} (\alpha')^{\frac{p-1}{2}} f_p \ecq
  f_p \in \bZ \ed
  \label{FluxQuant}
\end{align}
The total magnetic charge of the various sources must cancel on a compact manifold.
The resulting conditions are called the tadpole cancellation conditions, and are derived in section \ref{TC}.

From now on we will work in string units, setting $\alpha'=1$.

\subsection{Effective Low-Energy Potential}

Let us write down the effective 4D potential of supergravity in Einstein frame, where the action includes a canonical Einstein-Hilbert term. Instead of working directly with the dilaton, we will use the field
\begin{align}
  g = \frac{e^{\phi}}{\sqrt{V}} \ec
  \label{g}
\end{align}
where $V$ is the volume of the internal manifold.
To go to Einstein frame, make a field redefinition
\begin{align}
  \Gfour_{\mu\nu} = g^2 \GfourE_{\mu\nu} \ed
\end{align}
Since we are only interested in the potential, we will assume that the fields are all constant in spacetime. The relevant term in the 10D action is then
\begin{align}
  \frac{1}{2\kten^2} \int d^{10}x \sqrt{-\Gten} e^{-2\phi} \Rfour
  &= \frac{1}{2\kten^2} \int d^{4}x \sqrt{-\GfourE} \RfourE \ec
\end{align}
where $\RfourE$ is the Ricci scalar of $\GfourE$.
The 4D Planck scale $\Mp$ is $\Mp = (8\pi G)^{-1/2} = \kten^{-1}$.
Note that there is no controllable separation of scales between the 4D Planck scale and the string scale. 
This is because we chose to rescale the metric by the full $g$ field, and not just by its fluctuations around the background value. 
In this way we do not need to consider the stabilized value of $g$ (which may not exist).\footnote{
This is enough for our purposes, though it does mean that the 4D Planck scale we defined will not be the physical one in a particular solution.
In order to get quantitative results for a given vacuum, one would therefore have to perform an additional field redefinition.
}

Our next step is to define the 4D effective potential by dimensionally reducing on the internal manifold.
For a twisted torus background, the natural way to reduce is using the Scherk-Schwarz procedure \cite{Scherk:1979zr,Kaloper:1999yr}.
This reduction is consistent when the group manifold is unimodular \cite{Bergshoeff:2003ri}, namely when the structure constants that define the twisted torus obey
$f\ind{_A_B^B} = 0$; this is indeed the case for the manifolds we are considering.

In Scherk-Schwarz reductions the lowest `KK modes' depend on the internal coordinates, but only through the global frame \eqref{frame}.
For example, the $B_2$ field is reduced as
\begin{align}
  B_{\mu\nu}(x) dx^\mu \wedge dx^\nu +
  B_{\mu A}(x) dx^\mu \wedge \theta^A +
  B_{AB}(x) \theta^A \wedge \theta^B \ed
\end{align}
In 4D, $B_{\mu\nu}$ is a 2-form gauge field, $B_{\mu A}$ are gauge bosons, and $B_{AB}$ are scalars, or moduli.\footnote{Here we have set to zero the `KK gauge bosons' $A^A_\mu$ that arise from the metric. To include them, shift $\theta^A$ by $-A^A$.}
In this work we will focus on the lowest KK modes of the moduli, setting the 4D gauge fields to zero.
By using this prescription along with the background metric \eqref{g6}, the integral over the internal 6-manifold in the action becomes as trivial to compute as it is in the case of ordinary KK reductions.
For example, consider the $|H_3|^2$ term in \eqref{Skin}. 
Writing in in terms of global frame coordinates shows that it is constant on the internal manifold, and we have
\begin{align}
  - \frac{1}{4\kten^2} \int \! d^{10} x \, \sqrt{-\Gten} e^{-2\phi} |H_3|^2 &=
  - \frac{1}{4\kten^2} \int \! d^4 x \,
  \sqrt{-\GfourE} g^4 e^{-2\phi} 
  H_{A B C} H^{A B C} V
  \notag \\
  &= 
  - \frac{1}{4\kten^2} \int \! d^4 x \, \sqrt{-\GfourE}
  g^2 |H_3|^2
  \ed
\end{align}
Here we have ignored contributions from the $B_{\mu\nu}$ field for simplicity.

Now, define the 4D effective potential $U$ by\footnote{This differs by a factor of 2 from \cite{DeWolfe:2005uu}.}
\begin{align}
  S = \frac{1}{2\kten^2} \int d^4x \sqrt{-\GfourE} (-U) \ed
  \label{Ueff}
\end{align}
The pure supergravity (or `bulk') part of the effective potential, coming from the action \eqref{Skin}, is given by
\begin{align}
  U_{\rm bulk} = - g^2 R + \frac{1}{2} g^2 |H_3|^2
  + g^4 V \left( |\tilde{F}_2|^2 + |\tilde{F}_4|^2 + m_0^2 \right) \ed
  \label{Ukin}
\end{align}
The contribution of a space-filling D-brane or orientifold plane is
\begin{align}
  U_{\Dp} &= -2\kten^2 \mu_p g^3 \frac{V_{\Dp}}{\sqrt{V}}
  = (2\pi)^{7-p} g^3 \frac{V_{\Dp}}{\sqrt{V}} \ec 
  \label{UDp} \\
  U_{\Op} &= -2\kten^2 \tmu_p g^3 \frac{V_{\Op}}{\sqrt{V}}
  = - 4 \pi^{7-p} g^3 \frac{V_{\Op}}{\sqrt{V}} \ed
  \label{UOp}
\end{align}

In this work we will also consider spacetime-filling Kaluza-Klein (KK) 5-branes \cite{Gross:1983hb}.
These are metric objects that are charged under an isometry direction,
called the Taub-NUT cycle or charge cycle. 
Their effective action is given by
\begin{align}
  S_{\KK5} = - \int \! d^6\zeta \, \tau_{\KK5} \sqrt{-G_{\rm p.b.}}
  \ed
\end{align}
The tension is 
$\tau_{\KK5} = g_s^{-2} L_{\KKfive}^2 / 2\kten^2$,
where $L_{\KKfive}$ is the length of the Taub-NUT cycle.
\cite{Giveon:2008zn}.\footnote{
One can derive this from the tension of a D6 brane:
a KK5 monopole is lifted in M-theory to a KK6 monopole,
and dimensionally reducing on the KK6 charge cycle gives a D6 brane.
}
The contribution of such a brane to the effective potential is
\begin{align}
  U_{\KKfive} = \frac{L_{\KKfive}^2 V_{\KKfive}}{V} g^2 \ec
  \label{UKK}
\end{align}
where $V_{\KKfive}$ is the volume of the internal volume-minimizing 2-cycle that the brane wraps.

These are all the ingredients we will need in what follows.

\section{No-Go Result for Moduli Stabilization}
\label{nogo}

In this section we derive a no-go result for moduli stabilization of massive Type IIA supergravity on twisted tori, of the type presented in section \ref{TTG}.\footnote{
In particular, we assume that the monodromy $M$ is upper-triangular, with ones on the diagonal.}
We show that on such manifolds one cannot obtain a Minkowski solution, or a solution with parametrically small cosmological constant.
We begin by reviewing the discussion of \cite{Silverstein:2007ac} regarding the effective potential.

The general form of the effective potential \eqref{Ueff} is
\begin{align}
  U(g,\mu) = a(\mu) g^2 + b(\mu) g^3 + c(\mu) g^4 \ec
  \label{U}
\end{align}
where $g$ was defined in \eqref{g} and $\mu$ stands for the remaining moduli (such as the volume $V$). 
For now let us fix the moduli $\mu$ to some values $\mu=\mu_0$ and consider the dependence on $g$ alone, setting $U(g) = U(g,\mu_0)$. 

In order to have a minimum with positive $U$, both $a$ and $c$ must be positive, while $b$ must be negative.
$c$ is always positive in the presence of RR flux, and when the compactification manifold is a twisted torus, which has negative curvature, $a$ is also positive.
$b$ can be made negative by including orientifold planes.

Define the determinant
\begin{align}
  \Delta \equiv \frac{4ac}{b^2} \ed
  \label{Delta}
\end{align}
The condition for a minimum at $g_0>0$ to exist is $\Delta(\mu_0) < \frac{9}{8}$.
The sign of the potential at the minimum is determined by
\begin{align}
  U(g_0) \ge 0 \quad\iff\quad \Delta(\mu_0) \ge 1 \ed
\end{align}
Therefore, to have a vacuum with non-negative cosmological constant, we must have
\begin{align}
  1 \le \Delta(\mu_0) < \frac{9}{8} \ed
  \label{DeltaRange}
\end{align}
A Minkowski vacuum exists when $\Delta(\mu_0)=1$.

It can be difficult in practice to stabilize the moduli such that $\Delta(\mu)$ has a value within this range.
In \cite{Silverstein:2007ac,Haque:2008jz} it was shown that the situation is simplified when considering vacua with parametrically small and positive cosmological constant.
In this case finding a vacuum is equivalent to finding a minimum of the determinant $\Delta$.
In the next section we present a derivation of this result.
Minimizing the determinant can be easier than minimizing the potential, in particular because the determinant may have a global minimum while for the potential we are only considering local minima.

\subsection{Supergravity Vacua with Small Cosmological Constant}
\label{smallcc}

Let us show that finding a vacuum of the potential \eqref{U} with a positive and tunably small cosmological constant is equivalent to finding a nearby minimum of the determinant \eqref{Delta}.
First, notice that the potential can be written as
\begin{align}
  U = a g^2 \left( 1 + \frac{bg}{2a} \right)^2 
  + (\Delta-1) \frac{b^2 g^4}{4a} \ed
  \label{U2}
\end{align}
If $\Delta(\mu)$ has a minimum at which $\Delta=1$, then there is a Minkowski vacuum at $g=-2a/b$. At this point all fields will be stabilized, because $U$ grows quadratically in all directions in field space.
Furthermore, if $\Delta(\mu)$ has a minimum at a value that is larger than but sufficiently close to 1, then we expect there to be a nearby de Sitter vacuum with small cosmological constant.

Let us assume that a Minkowski vacuum exists at $\mu=\mu_0$.
At the vacuum the dilaton is given by $g = -\left. \frac{2a}{b} \right|_{\mu_0}$, and $\Delta(\mu_0) = 1$.
Choose one modulous $L$ of $\mu$, which takes the value $L_0$ at the vacuum.
Let $g$ depend on $L$ as
\begin{align}
  g(L) = -\left.\frac{2a}{b}\right|_{L,\hat{\mu}_0} \ec
\end{align}
where $\hat{\mu}$ stands for all the $\mu$ moduli except for $L$.
This defines a line through the pseudo-moduli space, that passes through the vacuum solution when $L=L_0$.
The potential along this line is $U(L) = U(g(L),L,\hat{\mu}_0)$, and it obeys
\begin{align}
  \left.\frac{dU}{dL}\right|_{L=L_0} = 0 \ecq
  \left.\frac{d^2U}{dL^2}\right|_{L=L_0} > 0 \ed
  \label{Ustab}
\end{align}
Using \eqref{U2} this potential can be written as
$U(L) = (\Delta - 1) \frac{4a^3}{b^2}$.
Using \eqref{Ustab} and the fact that $\Delta(\mu_0)=1$, we find that 
\begin{align}
  \left.\frac{dU}{dL}\right|_{L=L_0} &= 
  \left. \frac{d\Delta}{dL} \frac{4a^3}{b^2} \right|_{L=L_0} \Hence
  \left. \frac{d\Delta}{dL} \right|_{L=L_0} = 0 \ed
\end{align}
Differentiating again, we get
\begin{align}
  \left. \frac{d^2\Delta}{dL^2} \right|_{L=L_0} > 0 \ed
\end{align}
Since $L$ is an arbitrary modulous in $\mu$, this shows that $\Delta$ has a minimum at $\mu_0$.

Next, consider a vacuum with a small cosmological constant, which can be either positive or negative.
At this vacuum the moduli take the values $\mu=\mu'_0$.
In a Minkowski vacuum we would have $\Delta=1$, and here we assume that $\Delta(\mu'_0)$ is parametrically close to 1.
Equivalently, if we define $\Delta(\mu) = 1 + \delta(\mu)$, then $\delta_0 \equiv \delta(\mu'_0)$ is parametrically small.
We further assume that, in the limit $\delta_0 \to 0$, there is a Minkowski vacuum where all the moduli are at finite expectation values, and all have positive masses. 

We now show that $\Delta$ is stabilized for sufficiently small $\delta_0$.
Let $g$ depend on the other moduli as
\begin{align}
  g(\mu) = 
  - \frac{3+\sqrt{1-8\delta_0}}{4(1+\delta_0)}
  \left. \frac{2a}{b} \right|_{\mu} 
  = (-1 + 2\delta_0 + o(\delta_0^3)) 
  \left. \frac{2a}{b} \right|_{\mu} 
  \ed
\end{align}
It is easy to check that $g(\mu'_0)$ is the vacuum value of $g$.
Let us again choose an arbitrary modulous $L$, stabilized now at $L_0'$, and consider the line $g(L)$, setting all the other moduli besides $L$ to their values in $\mu_0'$.
For values of $L$ that are sufficiently close to $L'_0$, $\delta(L)$ will be close to $\delta_0$. The potential \eqref{U2} can then be written as
\begin{align}
  U(L) = \delta \frac{4a^3}{b^2} + o(\delta_0^2) \ed
\end{align}
Using the relations \eqref{Ustab} for the stability of $U(L)$ at $L'_0$ we find from the first derivative that
\begin{align}
  \left. \frac{d\delta}{dL} \right|_{L=L'_0} = o(\delta_0) \ec
\end{align}
and the second second derivative gives
\begin{align}
  0 < \left. \frac{d^2U}{dL^2} \right|_{L=L'_0} =
  \left. \frac{d^2\delta}{dL^2} \frac{4a^3}{b^2} \right|_{L=L'_0}
  + o(\delta_0) \ed
\end{align}
By assumption, the curvature of the potential is not suppressed in the limit $\delta_0 \to 0$, and the coefficients $a$, $b$, and $c$ go to finite values. Therefore $d^2\delta/dL^2$ is not suppressed in this limit, while $d\delta/dL$ is. This implies that, for sufficiently small $\delta_0$, there is a minimum of $\delta$ (and therefore of $\Delta$) close to $\mu'_0$.

A similar argument shows the converse. Given a minimum of the determinant with parametrically small $\delta$ (and finite curvature), the first derivative of $U$ is suppressed in the limit while the second is not, and therefore there is a nearby minimum of the potential.

\subsection{A No-Go Result for Twisted Tori}

As we now show, in a large class of models where the internal manifold is a twisted torus, the determinant $\Delta$ cannot in fact be stabilized. This is because the length $L_z$ of the base circle is a runaway direction of $\Delta$.
Therefore, in such models there are no vacua with parametrically small (or vanishing) cosmological constant.

Let us write down the contributions of each object to the coefficients of the potential \eqref{U}.
$a(\mu)$ receives contributions from NS-sector objects, such as the curvature and the $H_3$ flux; D-branes and orientifold planes contribute to $b(\mu)$; and $c(\mu)$ contributions come from RR field strengths. 
Based on the pieces \eqref{Ukin},\eqref{UDp},\eqref{UOp},\eqref{UKK}, and on contributions from NS5-branes, we can write schematically
\begin{align}
  a(\mu) &=
  -R + |H_3|^2 + L_{\KKfive}^2 \frac{V_{\KKfive}}{V} 
  + \frac{V_{\NSfive}}{V}
  \ec \notag \\ 
  b(\mu) &= 
  - \frac{V_{\Op}}{\sqrt{V}} + \frac{V_{\Dp}}{\sqrt{V}}
  \ec \notag \\ 
  c(\mu) &= 
  V \left( |\tilde{F}_2|^2 + |\tilde{F}_4|^2 + m_0^2 \right)
  \ed
  \label{abc}
\end{align}

The models we are considering can include general $H_3$ and RR fluxes (including $m_0$ flux), D-branes, NS5 branes, and KK5 monopoles.
The metric on the twisted torus is of the form \eqref{g6}.
We restrict orientifold planes to those that wrap the $dz$ cycle, in order not to spoil the motion of the inflaton brane as it rolls around the torus, and we do not consider warping.

Let us focus on the moduli $g$ and $L_z$ while keeping fixed $L_a$, the off-diagonal metric moduli, and the remaining non-metric moduli, and let us consider the dependence of the coefficients $a,b,c$ \eqref{abc} on $L_z$.
Starting with the contributions to $a$, the curvature \eqref{R} and volume \eqref{V} scale as $R \sim L_z^{-2}$ and $V \sim L_z$. The flux contribution $|H_3|^2$ can scale as $g^{zz} = L_z^{-2}$ or as $L_z^0$, depending on whether the $H_3$ flux includes a $dz$ factor or not.

As for the volume of an NS5 brane, we must take into account the fact that such a brane wraps a volume-minimizing submanifold in its homology class.
If the brane wraps a cycle on the fibre, then its volume is independent of $L_z$.
If it wraps the base circle (times an orthogonal cycle) then its volume is proportional to $L_z$.
More generally, the brane may wrap a `diagonal' direction such as $dz + \theta^5$.
On general grounds we expect the volume to grow at most linearly with $L_z$, namely we expect to have $V_{\NSfive} = L_z v(L_z)$, where $v'\le 0$.\footnote{
For example, consider a 1-brane that wraps a diagonal cycle in the class that corresponds to $dz + \theta^5$.
We would like to find the embedding $z(t),\xi(t)$ that minimizes its length, given the conditions $z(0)=\xi^5(0)=0$, $z(1)=\xi^5(1)=1$.
Let us choose the worldsheet coordinate such that $z(t) = t$, and let us set the lengths $L_a=1$ for simplicity.
The length of the brane is given by minimizing
\begin{align}
  \ell(L_z) = \int_0^n dt \sqrt{L_z^2 + \dot{\xi}^a \dot{\xi}^b g_{ab}}
  = \int_0^n dt \sqrt{L_z^2 + \left| \gamma(z) \dot{\xi}\right|^2}
  \ec
\end{align}
where $n\in\bZ$ counts the windings around the base circle.
The length is minimized by first choosing $\dot{\xi}^5 = 1$, and then choosing $\dot{\xi}^1,\dots,\dot{\xi}^4$ so as to cancel the remaining components of the vector $\gamma(z) \dot{\xi}$.
This leads to a closed geodesic for certain values of $n$, which depend on the monodromy. The allowed values determine the correct quantization of the cycle in the integer homology.
The length is then simply $n\sqrt{L_z^2 + 1}$.
We expect a similar conclusion for a brane wrapping a 2-cycle.
}

Finally, for a KK5-monopole the length $L_{\KKfive}$ of its charge cycle is independent of $L_z$, because $\dho_z$ is not an isometry.
Its contribution therefore takes the same form as that of an NS5-brane.
We find that
\begin{align}
  a(L_z) &= a_0 + \frac{a_1}{L_z} 
  + a_2 \tilde{v}(L_z)
  + \frac{a_3}{L_z^2} 
  \ec
  \label{a}
\end{align}
where $\tilde{v}(L_z)$ includes the contributions from both NS5 and KK5 branes, and obeys $\tilde{v}'\le 0$.
Notice that $a_0,a_1,a_2 \ge 0$, and $a_3 > 0$ (all contributions to $a_3$ are non-negative, and the curvature contribution is positive).

Next, $b$ can be written as
\begin{align}
  b(L_z) &= - b_0 L_z^{1/2} \ec
  \label{b}
\end{align}
because both O-planes and D-branes wrap $dz$ by assumption.
Since $a$ is positive, $b$ must be negative in order to have a minimum of the potential at finite $g$. Therefore $b_0>0$.
Finally, using similar arguments $c$ is given by
\begin{align}
  c(L_z) &= c_0 L_z + \frac{c_1}{L_z} \ec
  \label{c}
\end{align}
with $c_0,c_1 \ge 0$. 
It is now easy to verify that the determinant 
$\Delta(L_z) = 4ac / b^2$
is a monotonously decreasing function of $L_z$, and therefore cannot be stabilized. This completes the argument.

One may be able to evade the no-go result, for example by including D-branes or orientifold planes that do not wrap $dz$.
They will contribute a term $b_1 L_z^{-1/2}$ to $b$, and this can help stabilize $\Delta$ at large $L_z$.
In the context of inflation, one needs to make sure that these localized sources are sufficiently separated from the inflaton brane in order not to ruin its power-law potential.
Backreaction effects may also help stabilize the determinant, for example by introducing a term with a positive power in $\Delta(L_z)$ (see \cite{Dong:2010in} for a related discussion of backreaction effects).
There may also be vacua with small $L_z$, for which the supergravity approximation is not valid, and such vacua may also evade the no-go.

Finally, let us consider the model suggested in \cite{Silverstein:2007ac}, involving massive Type IIA on a product $\cM \times \tilde{\cM}$ of two twisted 3-tori with the same monodromy. This theory is orientifolded by $(z,\xi^1,\xi^2) \leftrightarrow (\tz,\txi^1,\txi^2)$, leading to O6-planes.
It is straightforward to verify explicitly that the no-go result applies to this model.
Let us write the metric as
\begin{align}
  \gint = L_z^2 (dz^2 + d\tz^2) + 2 g_{z\tz} dz d\tz
  + \sum_{a=1}^2 L_a^2 \left( \theta^a(z) + \theta^a(\tz) \right) \ec
\end{align}
where we have included the effect of the orientifold by setting $g_{AB} = g_{\tilde{A}\tilde{B}}$.
One can check that the modulous $L_+ = \sqrt{2(L_z^2 + g_{z\tz})}$ is a runaway direction of the determinant.

\section{Discrete Wilson Lines and Stabilization}
\label{DWL}

In this section we aim to clarify the role of discrete Wilson lines in moduli stabilization of supergravity theories.
We will show that the supergravity gauge transformations lead to obstructions when trying to turn on discrete Wilson lines. 

The effective potential \eqref{Ukin} of Type IIA supergravity includes Chern-Simons-like terms in which the gauge fields $C_1$ and $B_2$ appear without derivatives (the latter appears only when $m_0 \ne 0$).
To put it another way, the effective potential depends both on the `field strengths' $dC_1,dB_2$, and on the Wilson lines.
The Chern-Simons terms are useful in moduli stabilization, because they give masses to some of the Wilson line degrees of freedom of $C_1$ and $B_2$.
In this note we will only discuss the Chern-Simons-like terms in the potential that come from field strengths; we will not consider those terms that come from the Chern-Simons part of the supergravity action.

Discrete Wilson lines are related to torsion cycles; these are cycles that are not contractible by themselves, but become contractible when a certain number of them are taken together.
Recall that the homology groups $H_k(\cM;\bZ)$ of a closed manifold $\cM$ 
can be written as
\begin{align}
  H_k(\cM;\bZ) \cong \bZ \oplus \cdots \oplus \bZ \oplus T_k
  \ec \label{Hk}
\end{align}
where $T_k = \bZ_{N_1} \oplus \cdots \oplus \bZ_{N_i}$ is the torsion subgroup.
As we show in appendix \ref{secH1Z}, twisted tori generally admit torsion classes, whose ranks depend on the monodromy.

A Wilson line that is placed on a $\bZ_N$ cycle is discretized, and does not correspond to a dynamical degree of freedom. 
To see this, suppose $\Sigma$ is a $\bZ_N$ cycle, so that $N \Sigma \cong 0$ in homology. 
The Wilson line of a gauge field $A$ on $\Sigma$ obeys (schematically)
\begin{align}
  \left( e^{i \int_\Sigma A} \right)^N = 
  e^{i \int_{N\Sigma} A} = 1 \ed
  \label{disc}
\end{align}
The holonomy $\int_\Sigma A$ is therefore quantized in units of $2\pi/N$.

It will be important in what follows that a discrete Wilson line never corresponds to a globally defined form.
Rather, it is always a flat connection on a \emph{non-trivial} bundle, and is given only locally by a closed form.
Indeed, if $A$ is a globally defined closed form, then its integral on $\Sigma$ vanishes due to Stokes' theorem.
Such forms always correspond to continuous Wilson lines.

\subsection{General Manifolds}
\label{gc}

The presence of Chern-Simons terms in the action leads one to expect that discrete Wilson lines can affect the field strengths and the effective potential.
Therefore, as noted in \cite{Silverstein:2007ac}, discrete Wilson lines may play a role in moduli stabilization that is similar to that of fluxes.

Let us first see which contributions we can expect on general grounds, without restricting to a specific manifold.
Consider massive Type IIA supergravity on a closed, orientable manifold.
The field strengths
\begin{align}
  H_3 &= dB_2 \ec \notag \\
  \tilde{F}_2 &= dC_1 + m_0 B_2 \ec \notag \\
  \tilde{F}_4 &= dC_3 - C_1 \wedge H_3
  - \frac{m_0}{2} B_2 \wedge B_2 \ec
\end{align}
are gauge invariant, so they are given by globally defined forms on the manifold.
This property must be preserved if we turn on a non-global configuration of $C_1$ or $B_2$, such as a discrete Wilson line or flux.
This restriction leads immediately to the following conclusions.

If $m_0 \neq 0$, then a non-globally-well-defined contribution of $B_2$ to $F_2$ must be canceled by a similar contribution from $dC_1$.
If $B_2$ is a discrete Wilson line (namely if it is closed) then this cancellation is possible, at least locally.
Therefore, we expect that discrete Wilson lines can never contribute to $\tilde{F}_2$.
Otherwise, if $B_2$ generates flux, then its contribution cannot be canceled.
This leads to the tadpole cancellation condition $m_0 \int H_3 = 0$.

Next, consider contributions to $\tilde{F}_4$.
If the term $C_1 \wedge H_3$ is not globally defined then it must be canceled by turning on $C_3$.
If $C_1$ generates $\tilde{F}_2$ flux then this cancellation is not possible in general, and this leads to the tadpole cancellation condition $\int \tilde{F}_2 \wedge H_3 = 0$.
The cancellation is possible (at least locally) if $C_1$ is a discrete Wilson line.
These considerations do not completely rule out discrete $C_1$ contributions to $\tilde{F}_4$.
Indeed, if $C_1$ is a discrete Wilson line, $H_3$ is background flux, and $C_1 \wedge H_3$ is globally defined (even though $C_1$ is not), then it may contribute to $\tilde{F}_4$.
Contributions of discrete $B_2$ Wilson surfaces to $\tilde{F}_4$ are similarly restricted: they may be allowed only if $m_0 B_2 \wedge B_2$ is globally defined.\footnote{Notice that we have completely ruled out terms in $\tilde{F}_4$ that involve both a discrete Wilson line and a fluctuation of the background fields, because such terms are not globally defined for a general fluctuation.}

To summarize the discussion so far, we argued that discrete Wilson lines cannot make any contributions to $\tilde{F}_2$, and they might contribute to $\tilde{F}_4$ only when $C_1 \wedge H_3$ or $m_0 B_2 \wedge B_2$ are globally defined (where $H_3$ is quantized flux).
We see that discrete Wilson line contributions coincide with those of ordinary $\tilde{F}_4$ fluxes, and it raises the question of whether there is any difference between these two types of contributions.
The analysis below will suggest that these contributions can shift the integrals of $\tilde{F}_4$ by non-integer values.

\subsection{Twisted Tori}

In this section we write down the conditions for turning on discrete Wilson lines in Type IIA supergravity on twisted tori. 
The details appear in appendix \ref{DWLtori}.
The analysis is at the level of the gauge transformations, and does not take into account the higher-order gauge transformations that are present in $p$-form gauge theories, namely the transformations of the gauge parameter themselves.
A complete analysis that takes these transformations into account might reveal additional obstructions, and may modify the results we present here.

Consider IIA supergravity with $m_0=0$ on a twisted torus with monodromy $M$,
and with a general background configuration $C_1^{\bg}$, $B_2^{\bg}$, $C_3^{\bg}$ for the gauge fields.
If $B_2^{\bg}$ is a globally defined form, so there is no $H_3$ flux or discrete $B_2$ Wilson surfaces, then we can turn on a general $C_1$ Wilson line by setting
\begin{align}
  C_1 &= c_a d\xi^a \ecq
  C_3 = -B_2^{\bg} \wedge C_1
\end{align}
on the fundamental domain.
The solutions of $c_a M\ind{^a_b} = c_b$ correspond to continuous Wilson lines, while discrete Wilson lines obey
\begin{align}
  c_a M\ind{^a_b} - c_b \in \sqrt{2}\pi \bZ \ed
\end{align}
Turning on a discrete Wilson line in this way does not affect the field strengths.

If $B_2^{\bg}$ does generate $H_3$ flux, then we can turn on a discrete $C_1$ Wilson line on a $\bZ_N$ cycle only if $C_1 \wedge H_3$ is globally defined.
In appendix \ref{DWLtori} we show that the contribution of this piece to the 4-form flux is quantized according to
\begin{align}
  \sqrt{2} \int C_1 \wedge H_3 \in \frac{(2\pi)^3}{N} \bZ \ed
  \label{mquant}
\end{align}

Next, consider massive IIA. We can turn on $B_2$ Wilson surfaces by setting
\begin{align}
  B_2 = b_{ab} d\xi^a \wedge d\xi^b
\end{align}
on the fundamental domain. 
Solutions of $M^T b M = b$ correspond to globally defined forms, and therefore to continuous $B_2$ degrees of freedom, while discrete Wilson surfaces obey
\begin{align}
  \tilde{b}_{ab} \equiv (M^T b M - b)_{ab}  \in (2\pi)^2 \bZ \ed
\end{align}
Let us now assume that both $C_1^{\bg}$ and $B_2^{\bg}$ are globally defined.
By considering the effect of the $B_2$ gauge transformation on $C_1$ and $C_3$, one finds the following additional necessary conditions to turning on a discrete $B_2$,
\begin{gather}
  m_0 b_{ab} \in \sqrt{2}\pi \bZ
  \label{mbquant} \ec \\
  m_0 (M^T b M)_{ab} \tilde{b}_{cd} 
  d\xi^a \wedge d\xi^b \wedge d\xi^c \wedge d\xi^d = 0 \ed
\end{gather}
The first condition implies that a discrete $B_2$ Wilson surface obeys
\begin{align}
  m_0 \int B_2 \in \sqrt{2}\pi \bZ \ed
\end{align}
If $m_0 B_2 \wedge B_2$ is globally defined, then it will contribute to $\tilde{F}_4$ a piece $\frac{m_0}{2} B_2 \wedge B_2$. 
When $B_2$ is a sum of two discrete Wilson surfaces of ranks $N,M$, where $B_2 \wedge B_2$ is globally defined, then the flux contribution can take values in
\begin{align}
  \sqrt{2} \int \frac{m_0}{2} B_2 \wedge B_2 
  \in \frac{(2\pi)^3}{\gcd(N,M)} \bZ \ed
  \label{mquant2}
\end{align}

\section{Example: Twisted Torus with O6 Planes}
\label{expl}

In this section we consider the problem of moduli stabilization in Type IIA supergravity, when the internal manifold $\cM$ is a twisted torus with a specific monodromy.
We include an orientifold 6-plane that wraps the base circle, as well as KK5 branes and general fluxes.
We do not consider D-branes or orbifolds, which generally lead to additional moduli that may be difficult to stabilize.
We compute the general effective potential of the dilaton and the diagonal metric moduli $L_A$ with these ingredients.
We neglect the backreaction of the localized sources, as well as contributions from KK modes.

The monodromy of the twisted torus is $M=e^X$, where
\begin{align}
  X = \begin{pmatrix}
    0 & x_{12} &   &        &   \\
      & 0      &   &        &   \\
      &        & 0 & x_{34} & x_{35} \\
      &        &   & 0      & x_{45} \\
      &        &   &        & 0
  \end{pmatrix} \ed
  \label{X23}
\end{align}
The other components of $X$ vanish.
We demand that
\begin{align}
  x_{ij} \ne 0 \quad
  \mathrm{and} \quad
  x_{12}, x_{34}, x_{45}, 
  x_{35} + \frac{x_{34} x_{45}}{2} \in \bZ \ed
\end{align}
The last condition ensures that $M$ is an integer matrix.
On the manifold we have a global frame $\{\theta^A\}$ of 1-forms given by \eqref{frame}, and we take the metric to be \eqref{g6}
\begin{align}
  \gint
  &= L_z^2 (\theta^z)^2 + \sum_{a=1}^n L_a^2 (\theta^a)^2 \ed
\end{align}

\subsection{De Rham Cohomology}

In this section we write down the cohomologies with real coefficients of the twisted torus $\cM$.
The calculation is straightforward, and we illustrate it for the first cohomology. 

The global frame \eqref{frame} on $\cM$ is given by
\begin{align}
  \theta^z &= dz \ecq
  \theta^1 = d\xi^1 + x_{12} z d\xi^2 \ecq
  \theta^2 = d\xi^2 \ec
  \notag \\
  \theta^3 &= d\xi^3 + x_{34} z d\xi^4 + 
  \left( x_{35} z + \frac{x_{34} x_{45}}{2} z^2 \right) d\xi^5 \ecq
  \notag \\
  \theta^4 &= d\xi^4 + x_{45} z d\xi^5 \ecq
  \theta^5 = d\xi^5 \ed
  \label{frame23}
\end{align}
Differentiating these forms, we find
\begin{align}
  d\theta^1 &= x_{12} \theta^z \theta^2 \ec &
  d\theta^4 &= x_{45} \theta^z \theta^5 \ec \notag \\
  d\theta^2 &= 0 \ec \notag &
  d\theta^5 &= 0 \ec \\
  d\theta^3 &= \theta^z (x_{34} \theta^4 + x_{35} \theta^5) \ec &
  d\theta^z &= 0 \ed
  \label{d1}
\end{align}
Notice that a form that contains a $\theta^z$ factor is always closed, while a closed form that does not contain a $\theta^z$ factor is always non-trivial (non-exact).
We see that the first cohomology is
\begin{align}
  H^1(\cM,\bR) = \Span \{ \theta^z,\theta^2,\theta^5 \} \ed
  \label{cohom1}
\end{align}
Eqs. \eqref{d1} also tell us which 2-forms are exact.
To find the second cohomology, compute $d(\theta^a \theta^b)$ and identify which combinations are closed. Proceeding in this way, one can compute all the higher cohomologies, and here we write down the result in terms of harmonic representatives.
\begin{align}
  H^2(\cM,\bR) &= \Span \left\{ 
  \theta^z \theta^1 , \theta^z \theta^3 , 
  \theta^1 \theta^2 , \theta^2 \theta^5 ,
  \theta^4 \theta^5 , x_{45} \theta^1 \theta^5 - x_{12} \theta^2 \theta^4
  \right\} \ec
  \label{cohom2} \\
  H^3(\cM, \bR) &= \Span \Big\{
    \theta^z \theta^1 \theta^2 ,
    \theta^z \theta^1 \theta^3 ,
    \theta^z \theta^3 \theta^4 ,
    x_{34} \theta^z \theta^2 \theta^3 - x_{12} \theta^z \theta^1 \theta^4 ,
    \notag \\ 
    &\quad \qquad \quad \,
    \theta^1 \theta^2 \theta^5 ,
    \theta^2 \theta^4 \theta^5 ,
    \theta^3 \theta^4 \theta^5 ,
    x_{34} \theta^1 \theta^4 \theta^5 
    - x_{12} \theta^2 \theta^3 \theta^5
  \Big\} \ec
  \label{cohom3} \\
  H^4(\cM,\bR) &= \Span \Big\{
  \theta^z \theta^1 \theta^2 \theta^3 ,
  \theta^z \theta^1 \theta^3 \theta^4 ,
  \theta^z \theta^3 \theta^4 \theta^5 ,
  x_{45} \theta^z \theta^2 \theta^3 \theta^4
  - x_{12} \theta^z \theta^1 \theta^3 \theta^5 ,
  \notag \\
  &\quad \qquad \quad \,
  \theta^1 \theta^2 \theta^4 \theta^5 ,
  \theta^2 \theta^3 \theta^4 \theta^5
  \Big\}
  \ec \label{cohom4} \\
  H^5(\cM, \bR) &= \Span \Big\{
  \theta^z \theta^1 \theta^2 \theta^3 \theta^4 ,
  \theta^z \theta^1 \theta^3 \theta^4 \theta^5 ,
  \theta^1 \theta^2 \theta^3 \theta^4 \theta^5
  \Big\} \ec
  \label{cohom5} \\
  H^6(\cM, \bR) &= \Span \Big\{
  \theta^z \theta^1 \theta^2 \theta^3 \theta^4 \theta^5
  \Big\} \ed
  \label{cohom6}
\end{align}

\subsection{Orientifold Planes}

As explained in the introduction, it is useful to introduce orientifold planes in order to stabilize the dilaton with a positive cosmological constant.
We will divide by $\cO~=~\Omega_p (-1)^{F_L} T$ where $\Omega_p$ is worldsheet parity, $(-1)^{F_L}$ is the spacetime fermion number of left-moving modes, and $T$ is the following spacetime parity transformation.\footnote{In order not to ruin the inflaton brane's motion, we would like the orientifold plane to be extended in $dz$, so $T$ should not act on $z$.
It is easy to check that any orthogonal transformation $T$ that commutes with both $X$ and $L$ is a well-defined isometry on the manifold.}\textsuperscript{,}\footnote{
The factor $(-1)^{F_L}$ is included to ensure that $\cO^2 = 1$ \cite{Frey:2002hf}. Indeed, the parity transformation $T$ acts on a fermion $\psi$ as $\psi \to \Gamma \psi(T x)$ where $\Gamma = \gamma^3 \gamma^4 \gamma^5$, 
and $\Gamma^2 = -1$. Therefore $T^2 = (-1)^F$, where $(-1)^F$ is the spacetime fermion number. We now see that
\begin{align}
  \cO^2 
  = \Omega^2 T^2 (-1)^{F_L + F_R} 
  = (-1)^F (-1)^{F_L + F_R} = 1 \ed
\end{align}
}
\begin{gather}
  T : \quad (x^\mu, z, \xi^1, \xi^2, \xi^3, \xi^4, \xi^5) 
  \to (x^\mu, z, \xi^1, \xi^2, -\xi^3, -\xi^4, -\xi^5) \ed
\end{gather}
The projection introduces $N_{\Osix}=8$ $\Osix$-planes (counted in the fundamental domain). They are extended in spacetime and in the directions $z,\xi^1,\xi^2$, and are localized at $\xi^3,\xi^4,\xi^5\in \{0,\frac{1}{2}\}$.
(From now on we will use the separate notation $\alpha,\beta,...\!\in\!\{1,2\}$ and $i,j,...\!\in\!\{3,4,5\}$ for the two kinds of fibre directions.)

To see which field modes survive the orientifold, we need to work out the transformations of the bosonic fields under the orientifold action, and specifically their transformations under worldsheet parity $\Omega_p$.
As can be easily seen from the worldsheet action, the metric is $\Omega_p$-even while $B_2$ is $\Omega_p$-odd.

For the RR fields we can start with Type I theory which includes $C_2$, its dual $C_6$, and O9 planes \cite{polchinski1998string}. We then T-dualize in the directions $\xi^{3,4,5}$ to get to our $\Osix$-plane configuration, and see which RR field modes survive the orientifold.
For example, under this duality the Type I mode $C_{\mu\nu\rho345}$ maps to the mode $C_{\mu\nu\rho}$, which implies that the 3-form field $C_3$ has even worldsheet parity. By mapping the other modes one finds that the $p$-form gauge potential $C_p$ has worldsheet parity $(-1)^{(p+1)/2}$.
The worldsheet parities of the field strengths are
\begin{align}
  H_3(-) \ecq
  m_0(+) \ecq \tilde{F}_2(-) \ecq \tilde{F}_4(+) \ecq 
  \tilde{F}_6(-) \ecq \tilde{F}_8(+) \ecq \tilde{F}_{10}(-) \ed
  \label{fluxSigns}
\end{align}
The parity of $m_0$ can be determined from the relation $m_0 = *\tilde{F}_{10}$.

\subsection{Fluxes}
\label{TF}

The gauge-invariant field strengths $H_3$,$\tilde{F}_p$ can support fluxes, which are elements of the de Rham cohomology.
The fluxes that survive the orientifold projection, according to the signs \eqref{fluxSigns}, are
\begin{align}
  H_3 &= (2\pi)^2 
  \biggl[ 
  h_{z13} \theta^z \theta^1 \theta^3 +
  h_{125} \theta^1 \theta^2 \theta^5 +
  h_{345} \theta^3 \theta^4 \theta^5
  \notag \\ &\quad \qquad \quad 
  + h_{z23} \frac{\gcd(x_{12},x_{34})}{x_{12}^2 + x_{34}^2} \left(  
  x_{34} \theta^z \theta^2 \theta^3 
  - x_{12} \theta^z \theta^1 \theta^4
  \right)
  \biggr] 
  \ec \label{H3quant} \\
  \tilde{F}_2 &= \frac{2\pi}{\sqrt{2}} \biggl[ 
  f_{z3} \theta^z \theta^3 +
  f_{25} \theta^2 \theta^5 +
  \frac{f_{15}}{\gcd(x_{12},x_{45})} \left( x_{45} \theta^1 \theta^5 
  - x_{12} \theta^2 \theta^4 \right)
  \biggr] 
  \ec \label{F2quant} \\
  \tilde{F}_4 &= \frac{(2\pi)^3}{\sqrt{2}} \biggl[ 
  f_{z134} \theta^z \theta^1 \theta^3 \theta^4 
  + f_{1245} \theta^1 \theta^2 \theta^4 \theta^5 
  \notag \\ &\quad \qquad \quad \,\,
  + f_{z234} \frac{\gcd(x_{12},x_{45})}{x_{12}^2 + x_{45}^2} \left( 
  x_{45} \theta^z \theta^2 \theta^3 \theta^4 -
  x_{12} \theta^z \theta^1 \theta^3 \theta^5 
  \right)
  \biggr]
  \ec \label{F4quant} \\
  \tilde{F}_6 &= \frac{(2\pi)^5}{\sqrt{2}} f_6 
  \theta^z \theta^1 \theta^2 \theta^3 \theta^4 \theta^5 \label{F6quant} \ec \\
  m_0 &= \frac{1}{\sqrt{2} 2\pi} f_0 \label{m0quant} \ed
\end{align}
Note that turning on $\tilde{F}_6=*\tilde{F}_4$ is equivalent to turning on an $\tilde{F}_4$ flux that is polarized along the spacetime directions.

The flux quantization conditions \eqref{FluxQuant}
on the twisted torus (before taking the orientifold) imply that the coefficients above are all integers.
For example, if we integrate $H_3$ over the cycle $\Sigma_{345}$ dual to $\theta^3\theta^4\theta^5$ (this is most easily done by embedding this cycle at $z=0$), we find 
\begin{align}
  \int_{\Sigma_{345}} H_3 = 
  (2\pi)^2 h_{345} \int_0^1 d\xi^3 d\xi^4 d\xi^5 = (2\pi)^2 h_{345}
  \Hence h_{345} \in \bZ \ed
\end{align}
To show the quantization of the $\tilde{F}_2 \sim \theta^z \theta^3$ flux, first notice that the cycle $\sigma_3$ that wraps the $\xi^3$ direction is invariant under the monodromy.
We can therefore construct a cycle that wraps both this direction and $z$,
and it is easy to check that $\int \theta^z \theta^3 = 1$ on this cycle.
The condition for the flux
$\tilde{F}_4 \sim 
x_{45} \theta^z \theta^2 \theta^3 \theta^4 -
x_{12} \theta^z \theta^1 \theta^3 \theta^5$
can be worked out in a similar way by constructing the dual cycle.
Alternatively, we can use Poincar\'e duality and compute $\int \tilde{F}_4 \wedge \omega_2$, where $\omega_2$ represents an element of the cohomology with integer coefficients. In our case the relevant choice is
\begin{align}
  \omega_2 = \frac{
  x_{45} \theta^1 \theta^5 - x_{12} \theta^2 \theta^4
  }{\gcd(x_{12},x_{45})} \ec
\end{align}
which indeed has an integer period on any cycle.

We mention in passing that, as explained in section \ref{DWL}, turning on a discrete $B_2$ Wilson surface 
such that $-\frac{m}{2} B_2 \wedge B_2$ is a globally defined form can lead to a non-integer shift to the $\tilde{F}_4$ flux.
In our case we can shift the $f_{1245}$ flux by $\gcd^{-1}(x_{12},x_{45})$ (cf. \eqref{mquant2}) by turning on
\begin{align}
  B_2 = (2\pi)^2 \left( \frac{q}{x_{12}} d\xi^1 \wedge d\xi^5
  + \frac{r}{x_{45}} d\xi^2 \wedge d\xi^4 \right) \ecq
  q,r\in \bZ \ed
\end{align}
Notice that this configuration survives the orientifold projection.
The smallest shift we can achieve depends on $f_0$, which must satisfy the additional conditions
\begin{align}
  \frac{f_0 q}{x_{12}} \,,\, \frac{f_0 r}{x_{45}} \in \bZ \ed
\end{align}
For simplicity we will therefore keep $f_0$ general and $f_{1245}$ an integer.

This is the situation on the cover space.
As explained in \cite{Frey:2002hf}, after taking the orientifold we have additional (possibly unorientable) cycles, such as
\begin{align}
  z=\xi^1=\xi^2=0 \ecq
  0 \le \xi^3,\xi^4 < 1 \ecq
  0 \le \xi^5 < \frac{1}{2} \ed
  \label{Ocycle}
\end{align}
Integrating $H_3$ over this cycle leads to the stronger condition $h_{345} \in 2\bZ$.
In \cite{Frey:2002hf} it was shown that we can evade this restriction, by including `half-unit' flux contributions that are localized at the orientifold planes.
This allows us to turn on the same fluxes on the orientifold as we do on the cover space.

Note, however, that such localized flux contributions are proportional to delta functions, and lead to singular contributions to the supergravity action, $\int |H_3|^2 \sim \int \delta^2$.
Therefore, for the purpose of computing the effective potential within the supergravity approximation, we must use the more restrictive conditions that come from integrating over the halved orientifold space. These are
\begin{align}
  h_{ABC},f_{AB},f_{ABCD},f_6 \in 2\bZ \ecq
  f_0 \in \bZ \ed
  \label{finalQuant}
\end{align}

\subsection{Tadpole Cancellation}

As explained in appendix \ref{IIA}, fluxes must obey tadpole cancellation conditions. This is simply the statement that the total charge (coming from D-branes, O-planes, and fluxes) must cancel on a compact manifold.
In our case the $\Osix$-planes generate a tadpole for the $C_7$ field, and we will cancel this tadpole by turning on fluxes.
(We could also introduce $\Dsix$-branes to cancel part of the charge, but they would introduce additional moduli that may be difficult to stabilize.)
The relevant part of the O6-plane action \eqref{SDp} is
\begin{align}
  \sqrt{2} \tmu_p \int_{\wv} C_7 = 
  \sqrt{2} \tmu_p \int C_7 \wedge * J_7 \ec
\end{align}
where $J_7$ is the orientifold plane's charge density. In our case,
\begin{align}
  *J_7 = \sum_{\xi^i_{(0)} \in \{0,\frac{1}{2}\}}
  \delta^3\!\left( \xi^i - \xi^i_{(0)} \right) 
  \theta^3 \theta^4 \theta^5 \ed
\end{align}
The $C_7$ equation of motion reads (see \eqref{bianchi2})
\begin{align}
  d\tilde{F}_2 + m_0 H_3 + \sqrt{2} \tmu_6 \kten^2 *J_7 \ed
\end{align}
Integrating over any cycle, we find the condition
\begin{align}
  m_0 \int H_3 + \sqrt{2} \tmu_6 \kten^2 \int *J_7 = 0 \ed
\end{align}
If we perform the integral over the cycle \eqref{Ocycle} that is defined on the quotient space, we find the condition
\begin{align}
  f_0 h_{345} = 4 N_{\Osix} = 32 \ed \label{tadpole}
\end{align}
If we integrate over any other orthogonal cycle, we will find that $\int H_3=0$, and therefore we let
\begin{align}
  H_3 = (2\pi)^2 h_{345} \theta^3 \theta^4 \theta^5 \ed
\end{align}
The remaining tadpole conditions \eqref{brane-tadpole} are of the form $\int H_3 \wedge \tilde{F}_p = 0$ for $p=2,4,6$. They do not lead to additional constraints because the allowed RR fluxes all vanish when multiplied by $H_3$.

\subsection{Kaluza-Klein Monopoles}

We introduce Kaluza-Klein 5-branes \cite{Giveon:2008zn} that are charged under an isometry direction, are extended in spacetime, and wrap a 2-cycle in the internal manifold. The effective potential for these objects is written in \eqref{UKK}.

To find the isometries, notice that the cover space of our manifold is a group manifold $G$, and therefore has at least 6 isometries.
Five of the Killing vectors are given by $\frac{\dho}{\dho \xi^a}$, and an additional one overlaps with $\frac{\dho}{\dho z}$.
The isometries of our manifold correspond to the subgroup of $G$ that commutes with the lattice group by which we are dividing.
The vectors that point in the fibre directions must be invariant under the monodromy, namely they should satisfy
\begin{align}
  \frac{\dho}{\dho \xi^a} = \frac{\dho}{\dho \xi^b} M\ind{^b_a} \ed
\end{align}
The additional vector is not invariant under translations in the fibre direction, and does not survive the projection.
Our manifold therefore has two isometries in general, given by the globally-defined Killing vectors $\frac{\dho}{\dho \xi^1}$ and $\frac{\dho}{\dho \xi^3}$.

It turns out that the corresponding cycles $\Sigma_1$ and $\Sigma_3$ (those that wind around the directions $\xi^1$ and $\xi^3$) are torsion classes: they each become trivial when we wind them several times. This can be seen by computing the integer homology $H_1(\cM,\bZ)$, which is written down for a general twisted torus in section \ref{secH1Z}. In our case the homology contains the relations
\begin{align}
  x_{12} \Sigma_1 \cong 0 \ecq
  x_{34} \Sigma_3 \cong 0 \ed
  \label{torsion}
\end{align}
A KK5 brane that wraps a torsion cycle is not a source of dynamical gauge bosons, because if we take enough copies of it then the cycle it wraps will become trivial.
We can therefore introduce such KK5 branes without having to worry about canceling their total charge on the compact manifold.

As discussed in \cite{Silverstein:2007ac}, there may be topological restrictions on the number of KK5 branes in such a setting;
we will not go into such details here.
We also mention in passing that one may similarly include NS5 branes that wrap a torsion 2-cycle, and such branes would again not act as sources of dynamical gauge bosons.

\subsubsection{Orientifold Projection}

A KK5 brane is T-dual to an NS5 brane, if we perform the T-duality in the direction of the charge cycle.
To see which KK5 branes survive the orientifold projection, we can work out the NS5 branes that survive in the dual frame.

The twisted tori we are considering have simple properties under T-duality.
For example, if we T-dualize our manifold (with the monodromy \eqref{X23}) along $\frac{\dho}{\dho \xi^3}$, the result is another twisted torus with modified monodromy,
\begin{align}
  X \mapsto X' = \begin{pmatrix}
    0 & x_{12} &   &        &   \\
      & 0      &   &        &   \\
      &        & 0 & 0      & 0      \\
      &        &   & 0      & x_{45} \\
      &        &   &        & 0
  \end{pmatrix} \ec
  \label{X23p}
\end{align}
and with some background $H_3$ flux.
This can be seen by applying Buscher's rules \cite{Buscher:1987qj} to our metric \eqref{g6}.\footnote{See the appendix of \cite{Kachru:2002sk} for a summary of the rules in our conventions.}
More generally, if the monodromy has an $n \times n$ upper-triangular block with ones on the diagonal, then T-duality on the first coordinate of the block `untwists' it, leading to a $1 \times 1$ block plus an $(n-1)\times(n-1)$ block in the dual monodromy.

After T-dualizing in $\frac{\dho}{\dho \xi^3}$ we have O7 planes that are localized in $\xi^4,\xi^5$. 
NS5 branes in this frame couple to a $B_6$ gauge field, which is the electric-magnetic dual of the $B_2$ field. 
Both fields have the same odd parity under worldsheet parity. 
Therefore, the surviving space-filling NS5 branes correspond to 2-cycles that are odd under $\xi^{4,5} \to - \xi^{4,5}$.
Denoting the global frame of the dual manifold by $\theta'^A$, 
the odd (non-torsion) 2-cycles correspond to
\begin{align}
  \theta'^z \theta'^4 \ec
  \frac{x_{45} \theta'^1 \theta'^5 - x_{12} \theta'^2 \theta'^4}
  {\gcd(x_{12},x_{45})} \ec
  \theta'^2 \theta'^5 \ec
  \theta'^3 \theta'^5 \ed
\end{align}
Now, wrap an NS5 brane around one of these cycles and T-dualize again along $\frac{\dho}{\dho \xi^3}$ to get a KK5 brane. 
The cycle $\theta'^3 \theta'^5$ cannot be used because it is not orthogonal to the isometry direction. 
The cycle $\theta'^z \theta'^4$ becomes trivial in the dual torus. 

We find that the surviving KK5 branes that wrap non-torsion cycles, with charge along $\frac{\dho}{\dho \xi^3}$, are those that wrap a combination of
$\theta^2 \theta^5$
and
$(x_{45} \theta^1 \theta^5 - x_{12} \theta^2 \theta^4) / \gcd(x_{12},x_{45})$.
If we choose to include such branes then their charge will have to be canceled by including also anti-branes, but in the effective potential we are only sensitive to the total number of branes plus anti-branes.
Notice that the cycles we have found are all the 2-cycles in the original cohomology \eqref{cohom2} that are odd under the orientifold $\xi^{3,4,5} \to - \xi^{3,4,5}$, except for $\theta^z \theta^3$.
We could also introduce KK5 branes that wrap torsion cycles that are odd under the orientifold, as was done in \cite{Silverstein:2007ac}.
As explained above, such branes do not generate a net charge, and therefore do not need to be accompanied by anti-branes.

Next, to study KK5 branes with charge along $\frac{\dho}{\dho \xi^1}$ we T-dualize along this direction. The $2 \times 2$ block in the monodromy becomes trivial, and we have O5-planes that are localized in $\xi^{1,3,4,5}$.
The odd 2-cycles in this dual manifold are $\theta'^z \theta'^1, \theta'^z \theta'^3, 
\theta'^1 \theta'^2, \theta'^2 \theta'^5$. The surviving KK5 branes in the original frame, with charge along $\frac{\dho}{\dho \xi^1}$, can therefore wrap a combination of $\theta^z \theta^3$ and $\theta^2 \theta^5$. We again dropped cycles that are not orthogonal to the charge direction.

\subsection{General Effective Potential}

Collecting our results, and using the potential contributions \eqref{Ukin}, \eqref{UOp}, \eqref{UKK}, we find that the general effective potential \eqref{Ueff} in this class of models is given by
\begin{align}
  U &= a g^2 + b g^3 + c g ^ 4 \ec \\
  a &= 
  \frac{x_{12}^2}{2} \frac{L_1^2}{L_z^2 L_2^2} +
  \frac{x_{34}^2}{2} \frac{L_3^2}{L_z^2 L_4^2} +
  \frac{x_{35}^2}{2} \frac{L_3^2}{L_z^2 L_5^2} +
  \frac{x_{45}^2}{2} \frac{L_4^2}{L_z^2 L_5^2} +
  + 8 \pi^4 h_3^2 \frac{1}{L_3^2 L_4^2 L_5^2}
  \notag \\ &\quad
  + N^{\KK}_{z3} \frac{L_1}{L_2 L_4 L_5}
  + N^{\KK}_{25} \frac{L_1}{L_z L_3 L_4}
  \notag \\ &\quad
  + \frac{\tilde{N}^{\KK}_{15}}{\left|\gcd(x_{12},x_{45})\right|} \left( 
  \left| x_{45} \right| \frac{L_3}{L_z L_2 L_4} +
  \left| x_{12} \right| \frac{L_3}{L_z L_1 L_5}
  \right)
  + \tilde{N}^{\KK}_{25} \frac{L_3}{L_z L_1 L_4}
  \ec \\
  b &= -4\pi N_{\Osix} \sqrt{\frac{L_z L_1 L_2}{L_3 L_4 L_5}}
  \ec \\
  c &= 2\pi^2 \left\{
  f_{z3}^2 \frac{L_1 L_2 L_4 L_5}{L_z L_3}
  + f_{25}^2 \frac{L_z L_1 L_3 L_4}{L_2 L_5}
  + \frac{f_{15}^2}{\gcd^2(x_{12},x_{45})} \left( 
  x_{45}^2 \frac{L_z L_2 L_3 L_4}{L_1 L_5} + 
  x_{12}^2 \frac{L_z L_1 L_3 L_5}{L_2 L_4}
  \right)
  \right\}
  \notag \\ &\quad
  + 2^5 \pi^6 \Biggl\{
  f_{z134}^2 \frac{L_2 L_5}{L_z L_1 L_3 L_4}
  + \frac{f_{z234}^2 \gcd^2(x_{12},x_{45})}{(x_{12}^2 + x_{45}^2)^2} \left( 
  x_{45}^2 \frac{L_1 L_5}{L_z L_2 L_3 L_4} +
  x_{12}^2 \frac{L_2 L_4}{L_z L_1 L_3 L_5}
  \right)
  \notag \\ &\quad \qquad \qquad
  + f_{1245}^2 \frac{L_z L_3}{L_1 L_2 L_4 L_5}
  \Biggr\}
  \notag \\ &\quad
  + 2^9 \pi^{10} f_6^2 \frac{1}{L_z L_1 L_2 L_3 L_4 L_5}
  + \frac{f_0^2}{8\pi^2} L_z L_1 L_2 L_3 L_4 L_5 
  \ed
\end{align}
Here, the number of O6 planes is $N_{\Osix}=8$.
The coefficients $f_0$ and $h_3$ are related by \eqref{tadpole}, and the flux quantization conditions are given by \eqref{finalQuant}. 

$N^{\KK}$ and $\tilde{N}^{\KK}$ count KK5 branes (including anti-branes) with charge along $\frac{\dho}{\dho \xi^1}$ and $\frac{\dho}{\dho \xi^3}$, respectively.
The subscripts denote the cycle each brane wraps, in obvious notation. 
Those branes that do not wind around the base circle are localized at $z=0$.
All these branes either wrap a cycle on the fibre, or wrap the base circle and an orthogonal direction, and therefore their volumes (in the volume-minimizing embedding) are trivial to compute.
We did not include the contributions from the additional moduli that the branes introduce.

It is easy to see that the determinant $\Delta=4ac/b^2$ runs away at large $L_z$, which is consistent with our conclusion from section \ref{nogo}.
This rules out vacua with parametrically small cosmological constant in this class of models, but does not rule out de Sitter vacua in general.
We have searched for such vacua, but did not find any.

\section{Discussion}

In this note we attempted to make progress toward the construction of de Sitter vacua and of large field inflation models in string theory.
We classified the inflaton power-law potentials that arise from letting branes roll around twisted torus backgrounds, and found a variety of such potentials with powers smaller than 2.
This classification may be used as a starting point for constructing new models of large field inflation, along the lines of \cite{Silverstein:2008sg}.

We showed that a large class of twisted torus compactifications, in which orientifold planes and D-branes all wrap the base circle, cannot lead to de Sitter solutions with parametrically small cosmological constant.
It will be interesting to try and construct solutions that evade this no-go result, for example by using orientifold planes that are orthogonal to the base circle.
(Perhaps the simplest option is to orientifold by $(z,\xi) \to (-z,O \xi)$ where $O$ is a linear reflection operator that satisfied $\{O,X\}=0$.)
When using such models as backgrounds for brane inflation, one must make sure that the image inflaton branes are sufficiently separated in the transverse directions, so as not to ruin the inflaton potential.
We hope that the detailed example of section \ref{expl} can serve as a useful starting point for such an investigation.

Finally, we considered the role that discrete Wilson lines can play in moduli stabilization on general manifolds, and in particular on twisted tori.
We found that such contributions are restricted by gauge invariance.
The allowed contributions are similar to $\tilde{F}_4$ flux, except that on twisted tori these contributions can take fractional values according to \eqref{mquant},\eqref{mquant2}.
This property may be useful in moduli stabilization, because it extends the allowed choices of discrete parameters.
It will be interesting to understand whether the same fractional values are allowed on general manifolds, and we plan to return to this question in the future \cite{unpub}.

\section*{Acknowledgments}

The author would like to thank Ofer Aharony and Eva Silverstein for collaborating on early stages of this work and for commenting on early versions of this note, as well as Micha Berkooz, Cyril Closset, Elias Kiritsis, Zohar Komargodski, Itamar Shamir, and Ran Yacoby for useful discussions.
The author would especially like to thank Ofer Aharony for suggesting the idea for this project, and for many helpful and illuminating discussions.

This work was supported in part by an Israel Science Foundation center for excellence grant, by the German-Israeli Foundation (GIF) for Scientific Research and Development, by the Minerva foundation with funding from the Federal German Ministry for Education and Research, and by the I-CORE program of the Planning and Budgeting Committee and the Israel Science Foundation (grant number 1937/12).

\appendix

\section{Twisted Torus Geometry}
\label{TTGapp}

In this appendix we continue the discussion of section \ref{TTG}, giving additional details on the geometry of twisted tori.

\subsection{Riemannian Geometry}
\label{riem}

The metric is given by \eqref{g6}. 
It is convenient to define a rescaled global frame $\ttheta^A \equiv L_A \theta^A$, which is a vielbein of the metric (the $\theta^A$ were defined in \eqref{frame}).
These forms obey a Cartan equation of the form \eqref{cartan}, with structure constants
$\tf\ind{_A_B^C} = \frac{L_C}{L_A L_B} f\ind{_A_B^C}$.
Their non-vanishing components are
\begin{align}
  \tf\ind{_b_z^a} = - \tf\ind{_z_b^a} = 
  \frac{L_a}{L_b L_z} X_{ab} \ed
\end{align}

The spin connection is given by
\begin{align}
  \omega\ind{^a_z} = -\omega\ind{^z_a} &=
  \frac{1}{2} (\tf\ind{_b_z^a} + \tf\ind{_a_z^b}) \ttheta^b \ecq
  \omega\ind{^a_b} = \frac{1}{2}
  (\tf\ind{_a_z^b} - \tf\ind{_b_z^a}) \ttheta^z \ed
\end{align}
The curvature 2-form 
$
  R\ind{^A_B} = \frac{1}{2} R\ind{^A_B_C_D} \ttheta^C \wedge \ttheta^D = d\omega{^A_B} + \omega\ind{^A_C} \wedge \omega\ind{^C_B}
$
has the following non-vanishing components.
\begin{align}
  R\ind{^z_a} = - R\ind{^a_z} &=
  \frac{1}{2} \left[ 
  \tf\ind{_c_z^b} (\tf\ind{_b_z^a} + \tf\ind{_a_z^b})
  + \frac{1}{2} 
  (\tf\ind{_c_z^b} + \tf\ind{_b_z^c})
  (\tf\ind{_a_z^b} - \tf\ind{_b_z^a})
  \right] \ttheta^c \wedge \ttheta^z \ec\\
  R\ind{^a_b} &= \frac{1}{8} \left[ 
    (\tf\ind{_a_z^d} + \tf\ind{_d_z^a})
    (\tf\ind{_b_z^c} + \tf\ind{_c_z^b}) - (c \leftrightarrow d)
  \right]
  \ttheta^c \wedge \ttheta^d \ed
\end{align}
So far we did not assume that $X$ is strictly upper-triangular.
With this assumption the Ricci scalar is given by
$R = R\ind{^A_B_A_B} = - \frac{1}{2} \tf\ind{_a_z^b} \tf\ind{_a_z^b}$,
which gives the curvature \eqref{R}.

\subsection{Homology with Integer Coefficients}
\label{secH1Z}

Let us compute the first homology $H_1(\cM,\bZ)$. 
The fundamental group $\pi_1(\cM)$ is generated by the identifications \eqref{ident} that define the manifold. Let us denote them by
\begin{align}
  t_a &: \xi^a \to \xi^a + 1 \ec \\
  t_z &: (z,\xi^a) \to (z-1,M\ind{^a_b} \xi^b) \ed
\end{align}
The group is non-abelian, with commutators $[g,h]=g^{-1} h^{-1} g h$ given by 
\begin{align}
  [t_a,t_b] &= 1 \ec \\
  [t_a,t_z] &= \Pi_{b=1}^n (t_b)^{(M^{-1}-1)\ind{^b_a}} \ed
\end{align}
The integer homology is an abelian group, and it can be computed by setting the commutators of the fundamental group to the identity. Let $\Sigma_A$ denote the cycles defined by the translations $t_A$. The homology $H_1(\cM,\bZ)$ is then generated by $\Sigma_A$, up to the relations
\begin{align}
  \sum_{b=1}^n (M^{-1} - 1)\ind{^b_a} \Sigma_b \cong 0 \ecq
  a=1,\dots,n \ed
  \label{H1Z}
\end{align}

As a simple example, consider a twisted 3-torus with monodromy
\begin{align}
  M = \begin{pmatrix}
    1 & N \\ 0 & 1
  \end{pmatrix} \ec
\end{align}
where the fibre has coordinates $\xi^1,\xi^2$.
On this manifold the cycle that wraps $d\xi^1$ is a torsion cycle of rank $N$.

\section{Massive Type IIA Supergravity}
\label{IIA}

In this section we review massive Type IIA supergravity \cite{Romans:1985tz}. 
Let us begin by placing this theory on an orientable, compact manifold, without including branes or fluxes.
In the ordinary formulation, the theory includes a metric $\Gten$, a dilaton $\phi$, a 2-form NS gauge field $B_2$ with field strength $H_3$, and RR $p$-form gauge fields $C_1$ and $C_3$.
There is also a mass parameter $m_0$.
The bosonic action is (we follow the conventions of \cite{DeWolfe:2005uu})
\begin{align}
  S_{\IIA} &= S_{\kin} + S_{\mathrm{CS}} \ec \notag \\
  S_{\kin} &= \frac{1}{2\kten^2}
  \int d^{10}x \sqrt{-\Gten} \left[ 
  e^{-2\phi} \left( \Rten + 4 (\dho_\mu \phi)^2 
  - \frac{1}{2} |H_3|^2 \right)
  - |\tilde{F}_2|^2
  - |\tilde{F}_4|^2
  - m_0^2
  \right] \ec 
  \notag \\
  S_{\mathrm{CS}} &= 
  - \frac{1}{2\kten^2} \int \left[
  B_2 \wedge dC_3 \wedge dC_3 
  - \frac{m_0}{3} B_2 \wedge B_2 \wedge B_2 \wedge dC_3
  + \frac{m_0^2}{20} B_2 \wedge B_2 \wedge B_2 \wedge B_2 \wedge B_2
  \right] \ec
  \label{SIIA}
\end{align}
where
\begin{align}
  H_3 &= dB_2 \ec \label{H3} \\
  \tilde{F}_2 &= dC_1 + m_0 B_2 \ec \label{F2} \\
  \tilde{F}_4 &= dC_3 - C_1 \wedge H_3
  - \frac{1}{2} m_0 B_2 \wedge B_2 \ec \label{F4}
\end{align}
and
\begin{gather}
  2\kten^2 = (2\pi)^7 \alpha'^4 \ecq
  |F_p|^2 = \frac{1}{p!} F_{\mu_1 \dots \mu_p} F^{\mu_1 \dots \mu_p} \ed
\end{gather}
The action $S_{\IIA}$ and the field strengths $H_3$, $\tilde{F}_2$, $\tilde{F}_4$ are invariant under the gauge transformations
\begin{align}
  \delta_1 B_2 &= d\lambda_1 \ecq
  \delta_1 C_1 = -m_0 \lambda_1 \ecq
  \delta_1 C_3 = m_0 B_2 \wedge \lambda_1 \ec
  \notag \\
  \delta_0 C_1 &= d \Lambda_0 \ecq
  \delta_0 C_3 = - B_2 \wedge d \Lambda_0 \ec
  \notag \\
  \delta_2 C_3 &= d \Lambda_2 \ed
  \label{gauge0}
\end{align}
The field strengths $H_3$ and $\tilde{F}_p$ are, therefore, the natural gauge-invariant objects that can support fluxes.

\subsection{The Dual Field Strengths}

Let us compute the dual field strengths $\tilde{F}_6 = *\tilde{F}_4$, $\tilde{F}_8 = *\tilde{F}_2$.
The Bianchi identities for the dual field strengths are given by the equations of motion of $C_3$ and $C_1$. 
The part of the action that contributes to these equations is\footnote{In our conventions \cite{Nakahara},
$F_p \wedge * F_p = d^{10}x \sqrt{-\Gten} |F_p|^2$.}
\begin{align}
  - \frac{1}{2\kten^2} \int &\Big[
  \tilde{F}_2 \wedge * \tilde{F_2} +
  \tilde{F}_4 \wedge * \tilde{F_4} +
  B_2 \wedge F_4 \wedge F_4 
  - \frac{m_0}{3} B_2 \wedge B_2 \wedge B_2 \wedge F_4 
  \Big] \ed
  \label{Sdual}
\end{align}
Varying the action with respect to $C_3$, 
we have 
\begin{align}
  \delta S 
  &= \frac{1}{\kten^2} \int \delta C_3 \wedge d \Big[ 
  * \tilde{F}_4
  + B_2 \wedge F_4
  - \frac{m_0}{6} B_2 \wedge B_2 \wedge B_2
  \Big]
  \ed
\end{align}
Notice that $d(B_2 \wedge F_4) = d(C_3 \wedge H_3)$, so the equation of motion can be written as
\begin{align}
  d \Big[ 
  * \tilde{F}_4
  + C_3 \wedge H_3
  - \frac{m_0}{6} B_2 \wedge B_2 \wedge B_2
  \Big] = 0 \ed
  \label{C3eq}
\end{align}
Its solution is
\begin{align}
  \tilde{F}_6 \equiv *\tilde{F_4} =
  dC_5 - C_3 \wedge H_3
  + \frac{m_0}{6} B_2 \wedge B_2 \wedge B_2 \ed
  \label{F6}
\end{align}

Next, vary the action \eqref{Sdual} with respect to $C_1$. 
\begin{align}
  \delta S 
  &= \frac{1}{\kten^2} \int \delta C_1 \wedge d \Big[ 
  * \tilde{F}_2 
  + C_5 \wedge H_3 
  + \frac{m_0}{24} B_2 \wedge B_2 \wedge B_2 \wedge B_2 
  \Big] \ed
\end{align}
Solving the equation of motion, we find
\begin{align}
  \tilde{F}_8 \equiv * \tilde{F}_2 =
  dC_7 
  - C_5 \wedge H_3 
  - \frac{m_0}{24} B_2 \wedge B_2 \wedge B_2 \wedge B_2  \ed
  \label{F8}
\end{align}

\subsection{Democratic Formulation}

The democratic formulation of supergravity is a pseudo-action that involves $B_2$, the RR gauge fields, and their electric-magnetic duals \cite{Townsend:1995gp,Bergshoeff:2001pv}.
This pseudo-action can be used to derive the RR equations of motion and Bianchi identities of ordinary supergravity, but it includes an over-counting of degrees of freedom that must be removed by applying additional constraints.

In this section we write down the democratic formulation of massive Type IIA supergravity without any background fluxes. 
Our starting point are the field strengths \eqref{F2}, \eqref{F4} and their duals \eqref{F6}, \eqref{F8}. 
We also have $\tilde{F}_0 = -m_0$ and its dual $\tilde{F}_{10} = *\tilde{F}_0$. We can summarize these equations as
\begin{align}
  \tilde{F}_p = dC_{p-1} - C_{p-3} \wedge H_3
  - \frac{1}{(p/2)!} m_0 (-B_2)^{p/2} 
  \ecq p=0,2,4,6,8,10 \ed
  \label{Fp}
\end{align}
For $\tilde{F}_{10}$ this relation holds trivially: since it is a top form, the additional terms can be swallowed into $dC_9$, at least locally.
The field strengths are related by
\begin{align}
  *\tilde{F}_p = \tilde{F}_{10-p} \ed
  \label{dual-flux}
\end{align}
It is easy to check using \eqref{Fp},\eqref{dual-flux} that the field strengths satisfy the equations of motion (Bianchi identities)
\begin{align}
  d \tilde{F}_p + H_3 \wedge \tilde{F}_{p-2} = 0 \ec
  \label{bianchi}
\end{align}
where forms with negative rank vanish by definition.

We now write down the democratic formulation of the theory. Working in the algebra of forms (allowing the sum of forms with different ranks), let us define
\begin{align}
  C &= C_1 + C_3 + C_5 + C_7 + C_9 \ec \\
  \tilde{F} &= \tilde{F}_0 + \tilde{F}_2 + \tilde{F}_4 + \tilde{F}_6 + \tilde{F}_8 
  + \tilde{F}_{10} \ed
\end{align}
The field strengths \eqref{Fp} can now be written as
\begin{align}
  \tilde{F} = dC - C \wedge H_3 - m_0 e^{-B_2} \ed
  \label{demflux}
\end{align}
The field strengths are invariant under the gauge transformations $\delta_\lambda \tilde{F} = \delta_\Lambda \tilde{F} = 0$, where the gauge fields transform as
\begin{align}
  \delta_\lambda B_2 &= d\lambda_1 \ecq
  \delta_\lambda C = -m_0 e^{-B_2} \lambda_1 \ec
  \label{gauge1}
  \\
  \delta_\Lambda B_2 &= 0 \ecq
  \delta_\Lambda C = e^{-B_2} d\Lambda \ecq
  \Lambda = \sum_{n=0}^{n=4} \Lambda_{2n}
  \ed
  \label{gauge2}
\end{align}
These transformations generalize \eqref{gauge0}.
Let us define the democratic pseudo-action,
\begin{align}
  S_{\mathrm{dem.}} &= - \frac{1}{2\kten^2} \int \tilde{F} \wedge * \tilde{F}
  = - \frac{1}{2\kten^2} \int \sum_{n=0}^5 \tilde{F}_p \wedge * \tilde{F}_p \ed
  \label{Sdem}
\end{align}
This should be interpreted as an action of all the gauge potentials in $C$.
It is easy to see that this action reproduces the equations of motion \eqref{bianchi}, which can be written as
\begin{align}
  d \tilde{F} + H_3 \wedge \tilde{F} = 0 \label{dembianchi} \ed
\end{align}
The pseudo-action is not a full action because it over-counts the number of degrees of freedom. 
Indeed, after deriving the equations of motion we must reduce the number of degrees of freedom by imposing the constraints \eqref{dual-flux}, which we write as
\begin{align}
  *\tilde{F} = \tilde{F} \ed 
  \label{demdual}
\end{align}

\subsection{Fluxes and Tadpole Cancellation}
\label{TC}

$H_3$ and $\tilde{F}_p$ are the gauge-invariant field strengths, so they may support background fluxes.
The fluxes are quantized according to (setting $\ap=1$)
\begin{align}
  \int H_3 &= 2 \kten^2 \mu_5 h_3
  = (2\pi)^2 h_3 \ecq
  h_3 \in \bZ \ec
  \notag \\
  \sqrt{2} \int \tilde{F}_p &= 2 \kten^2 \mu_{8-p} f_p 
  = (2\pi)^{p-1} f_p \ecq
  f_p \in \bZ \ed
\end{align}
These conditions determine the normalization of the $U(1)$ gauge transformations \eqref{gauge1},\eqref{gauge2}.
Given gauge parameters $\lambda_1$, $\Lambda_{p}$, the corresponding group elements $g \in U(1)$ are given by
\begin{align}
  g(\lambda_1) = \exp \left( \frac{i}{2\pi} \int \lambda_1 \right)
  \ecq
  g(\Lambda_{p}) = \exp \left( 
  \frac{\sqrt{2} i}{(2\pi)^{p}}
  \int \Lambda_{p}
  \right) \ed
  \label{gaugeNorm}
\end{align}
These can be computed, for example, by explicitly constructing the non-trivial bundle that gives flux on a torus.

Not all combinations of quantized background fluxes are allowed.
As the Bianchi identities \eqref{bianchi} show, certain combinations of field strengths act as magnetic sources for other gauge fields. 
Such sources must cancel when integrated over compact cycles. 
Indeed, let us integrate the Bianchi identity \eqref{bianchi} for $\tilde{F}_{p+2}$ over a $(p+3)$-cycle, 
\begin{align}
  \int d \tilde{F}_{p+2} + \int H_3 \wedge \tilde{F}_{p} = 0 \ed
  \label{intbianchi}
\end{align}
Since $\tilde{F}_{p+2}$ is gauge-invariant under \eqref{gauge1},\eqref{gauge2}, it is a globally-defined form on the manifold. 
The first term in \eqref{intbianchi} therefore vanishes, and we are left with
\begin{align}
  \int H_3 \wedge \tilde{F}_{p} = 0 \ecq
  p=0,2,4,6
  \ed
  \label{tadpole1}
\end{align}
These are the so-called tadpole cancellation conditions in the absence of additional brane sources. 

When turning on fluxes the CS piece of the action \eqref{SIIA} receives certain corrections \cite{DeWolfe:2005uu}, while the kinetic piece is unchanged.
Fluxes can be added to the field strengths independently of the gauge fields, or they can be added to the gauge fields directly, resulting in gauge fields that are not globally-defined. 

Now let us introduce $p$-branes (these can be D-branes or O-planes) with coupling to the RR fields given by
\begin{align}
  S_p = \sqrt{2} \mu \int_{\mathrm{w.v.}} C_{p+1} + \cdots =
  \sqrt{2} \mu \int C_{p+1} \wedge * J_{p+1} + \cdots \ed
\end{align}
Here $J_{p+1}$ is the charge density.
The terms we are omitting involve RR gauge fields of lower rank. They will not be important for our purposes, though in general they can affect the Bianchi identities \cite{Douglas:1995bn},\cite{Green:1996bh}.
Adding this contribution to the action, the Bianchi identities \eqref{dembianchi} become 
\begin{align}
  d\tilde{F} + H_3 \wedge \tilde{F} + 
  \sqrt{2} \mu \kten^2 * J = 0 \ec
  \label{bianchi2}
\end{align}
where $J$ is the sum of charge densities for all the branes. Integrating this over a cycle, we find the tadpole cancellation conditions
\begin{align}
  \int H_3 \wedge \tilde{F} + 
  \sqrt{2} \mu \kten^2 \int * J = 0 \ed
  \label{brane-tadpole}
\end{align}

\section{Discrete Wilson Lines on Twisted Tori}
\label{DWLtori}

In this section we work out the detailed conditions for turning on discrete Wilson lines in Type IIA supergravity on twisted tori.
We illustrate the conditions via several examples.

\subsection{Flat $C_1$ Connections}
\label{C1DWL}

Consider massive IIA supergravity on a twisted torus with monodromy $M$, with field strengths \eqref{IIAH3},\eqref{IIAF2},\eqref{IIAF4}.
Start with a general background configuration $C_1^{\bg}$, $B_2^{\bg}$, $C_3^{\bg}$ for the gauge fields, assuming for now that $B_2^{\bg}$ is a globally defined form, so there is no $H_3$ flux.
Let us turn on a general Wilson line (flat connection) by setting $C_1^\tot = C_1^{\bg} + C_1$, where
\begin{align}
  C_1 = c_a d\xi^a + c_z dz
\end{align}
in the fundamental domain $z,\xi^a \in [0,1)$, and $c_a$,$c_z$ are real coefficients.
The map $f(z,\xi) = (z-1,M\xi)$ defines the twisted torus identification when going around the base circle.
Ordinary, continuous Wilson lines, which are flat connections on a trivial gauge bundle, are given by those forms that are invariant under $f$,
\begin{align}
  C_1 = f^* C_1
  \quad
  \Longleftrightarrow 
  \quad
  c_a d\xi^a = c_a M\ind{^a_b} d\xi^b \ed
\end{align}
It is easy to see that the space of continuous Wilson lines is spanned by $\left\{ dz , d\xi^a \,|\, d\xi^a = \theta^a \right\}$, and we set the corresponding coefficients to zero.
Discrete Wilson lines are flat connections that are periodic up to a non-trivial gauge transformation,
\begin{align}
  f^* C_1 - C_1
  = (c_a M\ind{^a_b} - c_b) d\xi^b = d\Lambda_0 \ed
  \label{C1pbtrans}
\end{align}
If we set the gauge parameter to $\Lambda_0 = (c_a M\ind{^a_b} - c_b) \xi^b$, it defines a gauge transformation $\exp(\sqrt{2} i\Lambda_0)$ \eqref{gaugeNorm} that is periodic on the fibre only if\footnote{The peculiar factors of $\sqrt{2}$ are due to our choice of conventions for the RR fields.}
\begin{align}
  c_a M\ind{^a_b} - c_b \in \sqrt{2}\pi \bZ \ed
  \label{C1DWLquant}
\end{align}
This is the quantization condition for discrete $C_1$ Wilson lines.
Discrete Wilson lines are also periodic, for the same reason as ordinary Wilson lines: we can shift $c_a$ by performing a gauge transformation $C_1 \to C_1 + \sqrt{2}\pi d\xi^a$ on the fundamental domain.

In an ordinary gauge theory this would be the end of the story, but in supergravity the gauge transformation \eqref{C1pbtrans} also affects $C_3$.
If we set the total $C_3$ to $C_3^\tot = C_3^{\bg} + C_3$, then the extra piece should transform as
\begin{align}
  f^* C_3 - C_3 = - B_2^{\bg} \wedge d\Lambda_0 \ed
  \label{C1lineC3t}
\end{align}
By assumption $B_2^{\bg}$ is globally defined, and therefore we can set
\begin{align}
  C_3 = - B_2^{\bg} \wedge C_1
  \label{C3forC1}
\end{align}
patch-by-patch on the bundle.
With this ansatz, we find that all the field strengths are unchanged. In particular, denoting by $H_3^{\bg}$ and $\tilde{F}_4^{\bg}$ the background field strengths, we see that
\begin{align}
  \tilde{F}_4^\tot - \tilde{F}_4^{\bg} = 
  dC_3 - C_1 \wedge H_3^{\bg} = 0 \ed
\end{align}
Therefore, in the absence of $H_3$ flux, discrete $C_1$ Wilson lines do not contribute to the effective potential.

If we introduce $H_3$ flux by adding a non-global piece $B_2^\flux$ to the background (assuming $m_0=0$), then there are two possibilities.
If $B_2^\flux \wedge d\Lambda_0$ vanishes then $C_3$ requires no modification, and we do not encounter an obstruction to turning on $C_1$ and $H_3$ together.
This analysis is at the level of the gauge transformations.
$p$-form gauge theories have higher-order gauge transformations (namely, the gauge parameters themselves can undergo gauge transformations), and taking these into account may reveal an obstruction.
Analyzing these higher-order gauge transformations is beyond the scope of this work.
If there are no additional obstructions, then in this case $\tilde{F}_4$ will receive a contribution from the term $C_1 \wedge H_3$.
This situation is only possible if $C_1 \wedge H_3$ is globally defined (even though $C_1$ is not).

The other possibility is that $B_2^\flux \wedge d\Lambda_0 \ne 0$, in which case we will have to add to $C_3$ a piece $- B_2^\flux \wedge C_1$.
This piece will generally not be globally defined because of $B_2^\flux$, but we may be able to account for this by adding a gauge transformation $C_3 \to C_3 + d\Lambda_2$.
Here we may face an obstruction, because the gauge transformation defined by $\Lambda_2$ must be well-defined (periodic). This will generally lead to a quantization condition involving both the flux and the discrete Wilson line coefficients.
In this case, if we evade the obstruction, then we find that none of the field strengths are modified by the discrete Wilson line.
This must be the case if $C_1 \wedge H_3$ does not have a non-trivial globally defined part.

\subsection{$C_1$ Example}
\label{C1expl}

Let us illustrate these ideas with an example.
Consider the manifold $\cM \times T^3$, where $\cM$ a twisted 3-torus with monodromy
\begin{align}
  M = \begin{pmatrix}
    1 & N \\ 0 & 1
  \end{pmatrix}
\end{align}
and coordinates $z,\xi^1,\xi^2$, and the coordinates of the 3-torus are $\xi^3,\xi^4,\xi^5$, with periodicity 1.
On this manifold we can turn on a discrete $C_1$ Wilson line by setting $C_1 = \sqrt{2}\pi \frac{q}{N} d\xi^1$.
The quantization condition \eqref{C1DWLquant} reads $q \in \bZ$,
and the Wilson line is periodic under $q \to q+N$.
When $B_2^{\bg}$ is globally defined, we can set $C_3$ as in \eqref{C3forC1} and get a consistent construction for which all the field strengths vanish.

Now, suppose we turn on a flux 
\begin{align}
 H_3^{\bg} 
 = (2\pi)^2 h \, d\xi^2 \wedge d\xi^3 \wedge d\xi^4
 \ecq h \in \bZ
\end{align}
while setting $m_0=0$ to cancel the tadpole.
We choose the gauge 
$B_2^{\bg} = (2\pi)^2 h \, \xi^4 d\xi^2 \wedge d\xi^3$.
Notice that the transformation \eqref{C1lineC3t} of $C_3$ vanishes (here $\Lambda_0 = \sqrt{2}\pi q \xi^2$), so we can set $C_3=0$.
The field strength is then
\begin{align}
  \tilde{F}_4 = - C_1 \wedge H_3^{\bg} =
  - \frac{(2\pi)^3}{\sqrt{2}} \frac{qh}{N}
  d\xi^1 \wedge d\xi^2 \wedge d\xi^3 \wedge d\xi^4 \ed
\end{align}
As expected, this is a (non-trivial) globally defined form.
However, notice that the coefficient is not quantized according to the usual condition \eqref{FluxQuant}.
When $q \in N\bZ$ the $C_1$ Wilson line becomes trivial, and then the coefficient is quantized according to \eqref{FluxQuant}, so such a flux can be implemented by turning on $C_3$ as usual.
Overall, we see that in this example the field strength can take the discrete values
\begin{align}
  \sqrt{2} \tilde{F}_4 = 
  (2\pi)^3 \tilde{f} d\xi^1 \wedge d\xi^2 \wedge d\xi^3 \wedge d\xi^4
  \ecq
  \tilde{f} \in \frac{1}{N} \bZ \ed
\end{align}

Next, let us turn on a different flux
\begin{align}
 H_3 = (2\pi)^2 h \, d\xi^3 \wedge d\xi^4 \wedge d\xi^5
 \ecq h \in \bZ \ed
\end{align}
Now $C_1 \wedge H_3$ is not globally defined (it is not invariant under $f$), so we do not expect the discrete $C_1$ to contribute to $\tilde{F}_4$.
Choosing the gauge to be $B_2^\bg = (2\pi)^2 h \, \xi^3 d\xi^4 \wedge d\xi^5$, the transformation \eqref{C1lineC3t} does not not vanish and we set
\begin{align}
  C_3 = - B_2^\bg \wedge C_1 = c \, \xi^3 d\xi^1 \wedge d\xi^4 \wedge d\xi^5 
  \ecq
  c = - \frac{(2\pi)^3}{\sqrt{2}} \frac{qh}{N} \ed
\end{align}
This is periodic in $\xi^3$ up to a gauge transformation $C_3 \to C_3 + d\Lambda_2$, $\Lambda_2 = c \,\xi^1 d\xi^4 \wedge d\xi^5$.
This transformation is well defined (see \eqref{gaugeNorm}) only if
\begin{align}
  \frac{qh}{N} \in \bZ \ed
\end{align}
This condition is an obstruction to turning on both the $C_1$ Wilson line and the $H_3$ flux.
If this condition is obeyed, we find that all field strengths vanish as expected.

\subsection{Flat $B_2$ Connections}
\label{B2DWL}

Next, we consider discrete $B_2$ Wilson surfaces. 
For simplicity we will assume that the background $C_1^\bg$ is globally defined, while $B_2^{\bg}$ and $C_3^\bg$ may include flux potentials.
Let us turn on a general flat connection by setting $B_2^\tot = B_2^{\bg} + B_2$, with
\begin{align}
  B_2 = b_{ab} d\xi^a \wedge d\xi^b + b_{a} d\xi^a \wedge dz
\end{align}
in the fundamental domain.
$B_2$ should be invariant under $f$ up to a gauge transformation $B_2 \to B_2 + d\lambda_1$.

Consider first the terms that include a $dz$ factor.
They must be invariant under $f$ up to a gauge transformation, namely
\begin{align}
  \tilde{b}_a d\xi^a \wedge dz = d\lambda_1 \ecq
  \tilde{b}_a \equiv b_b M\ind{^b_a} - b_a \ed
  \label{lambda1}
\end{align}
Note that this gauge transformation occurs on a domain (an intersection of two patches on the bundle) that does not wrap $dz$, because it relates $B_2$ and $f^* B_2$.
The equation \eqref{lambda1} can be solved by setting $\lambda_1 = \tilde{b}_{a} d\xi^a z$ on this domain. The resulting gauge transformation, after integrating in some direction $d\xi^b$, is
\begin{align}
  \exp \left( \frac{i}{2\pi} \int \lambda_1 \right)
  = \exp \left( \frac{i (b_{c} - b_{a} M\ind{^a_c}) z }{2\pi} \right)
  \ed
\end{align}
This does not lead to a quantization condition (even though the bundle is not trivial), and therefore there are no discrete $B_2$ Wilson lines that involve $dz$. From now on we set $b_a = 0$.

The condition on $B_2$ is now
\begin{align}
  f^* B_2 - B_2 = \tilde{b}_{ab} d\xi^a \wedge d\xi^b = d\lambda_1 \ecq
  \tilde{b} \equiv M^T b M - b
  \ed
\end{align}
The continuous connections are those that obey $\tilde{b}_{ab} = 0$.
The discrete connections live on a non-trivial bundle with gauge parameter $\lambda_1 = \tilde{b}_{ab} \xi^a d\xi^b$.
Demanding periodicity of the gauge transformation 
$\exp\left( \frac{i}{2\pi} \int \lambda_1 \right)$ 
on the fibre leads to the quantization condition\footnote{
As we show below, it is possible that different terms in $B_2$ give proportional contributions to $d\lambda_1$.
In such a case one must quantize each term separately, because different terms correspond to different torsion classes.}
\begin{align}
  \tilde{b}_{ab} = b_{cd} M\ind{^c_a} M\ind{^d_b} - b_{ab} 
  \in (2\pi)^2 \bZ \ed
  \label{btildecond}
\end{align}
If $m_0$ vanishes then this defines a consistent gauge bundle, in which the $B_2$ piece does not affect the field strengths or the potential.\footnote{Here we use the fact that $C_1$ is globally defined. If it was not global, turning on $B_2$ would alter the gauge transformation $C_1 \to C_1 + d\Lambda_0$, $C_3 \to C_3 - B_2^\tot \wedge d\Lambda_0$, and could lead to an obstruction.}

Now let us assume that $m_0 \ne 0$, which implies that there is no $H_3$ flux due to tadpole cancellation, and therefore $B_2^{\bg}$ is globally defined.
In this case we generally have to introduce corrections $C_1^\tot = C_1^{\bg} + C_1$ and $C_3^\tot = C_3^{\bg} + C_3$ to account for the fact that the $B_2$ gauge transformation also affects the RR gauge fields through
\begin{align}
  f^* C_1 - C_1 &= -m_0 \lambda_1 \ec \notag \\
  f^* C_3 - C_3 &= m_0 B_2^{\tot} \wedge \lambda_1
  \label{fsC13} \ed
\end{align}
Notice that the $C_3$ transformation is well defined, in the sense that it is independent of the patch on which $B_2^\tot$ is evaluated, only if
\begin{align}
  m_0 d\lambda_1 \wedge \lambda_1 = 0 \ed \label{consist}
\end{align}
The $C_1$ gauge transformation in \eqref{fsC13} can be accommodated by setting
$C_1 = - m_0 b_{ab} \xi^a d\xi^b$.
It follows that $\tilde{F}_2$ receives no contribution from $B_2$, since $dC_1 = - m_0 B_2$.
This is in accordance with our general argument above.

Now, this configuration (on the fundamental domain) is periodic in the fibre direction up to additional gauge transformations
\begin{align}
  C_1 &\to C_1 + d\Lambda_0^{(a)} \ec \notag \\
  C_3 &\to C_3 - B_2^{\tot} \wedge d\Lambda_0^{(a)} \ec \notag \\
  \Lambda_0^{(a)} &= - m_0 b_{ab} \xi^b \ed
  \label{deltaC13}
\end{align}
These, in turn, lead to periodic gauge transformations $\exp(\sqrt{2} i \Lambda_0^{(a)})$ only if $m_0 b_{ab} \in \sqrt{2} \pi \bZ$. 
This condition can be written as
\begin{align}
  f_0 b_{ab} \in (2\pi)^2 \bZ \ec
  \label{fbobst}
\end{align}
where $\sqrt{2} m_0 = f_0 / 2\pi$, $f_0 \in \bZ$.
It follows that for a discrete $B_2$ Wilson surface,
\begin{align}
  f_0 \int B_2 \in (2\pi)^2 \bZ \ed
\end{align}

In order to complete the construction we must also account for the gauge transformations of $C_3$ under $\lambda_1$ and $\Lambda_0^{(a)}$.
The natural guess is to take $C_3 = - B_2^{\tot} \wedge C_1$.
This transforms correctly, as in \eqref{deltaC13}, when going around the fibre, but when going around the base circle we have
\begin{align}
  f^* C_3 - C_3 &= - B_2^{\bg} \wedge (f^* C_1 - C_1)
  - f^* B_2 \wedge f^* C_1 + B_2 \wedge C_1
  \notag \\
  &= m_0 B_2^{\bg} \wedge \lambda_1 - (f^* B_2 - B_2) \wedge f^* C_1
  - B_2 \wedge (f^* C_1 - C_1)
  \notag \\
  &= m_0 B_2^{\tot} \wedge \lambda_1
  - d\lambda_1 \wedge f^* C_1
  \ed
\end{align}
Compare this with the desired transformation \eqref{fsC13}, which does not include the second term in the last line.
If this term vanishes, namely if
\begin{align}
  d\lambda_1 \wedge f^* C_1 &=
  (f^* B_2 - B_2) \wedge f^* C_1 \notag \\ 
  &=
  -m_0 b_{a'b'} M\ind{^{a'}_a} M\ind{^{b'}_b} \tilde{b}_{cd} 
  \xi^a d\xi^b \wedge d\xi^c \wedge d\xi^d
  = 0 \ec
  \label{dlfC1}
\end{align}
then we are done.
Otherwise, if it is merely closed,
then we can try to remove it with an additional gauge transformation $C_3 \to C_3 + d\Lambda_2$.
Using $dC_1 = -m_0 B_2$, the condition of being closed is equivalent to
\begin{align}
  m_0 (f^* B_2 - B_2) \wedge f^* B_2 = 0 \ed
\end{align}
We now make use of the condition \eqref{consist}, which implies that
\begin{align}
  m_0 d\lambda_1 \wedge d\lambda_1 = 
  m_0 (f^* B_2 - B_2) \wedge (f^* B_2 - B_2) = 0 \ed
\end{align}
Combining the two conditions, we see that 
$m_0 f^* (B_2 \wedge B_2) = m_0 B_2 \wedge B_2$,
namely that $m_0 B_2 \wedge B_2$ is necessarily a globally defined form, as we anticipated.

To summarize, a discrete flat connection of the form $B_2 = b_{ab} d\xi^a \wedge d\xi^b$, subject to the quantization \eqref{btildecond} and in the absence of $H_3$ and $\tilde{F}_2$ flux, can be turned on if the conditions \eqref{fbobst} and \eqref{dlfC1} are satisfied. 
(If \eqref{dlfC1} is not satisfied but $d\lambda_1 \wedge f^* dC_1=0$, it may still be possible to turn the connection on, with an additional $C_3$ gauge transformation.)
The corrections to the RR gauge fields are given by
\begin{align}
  C_1 = -m_0 b_{ab} \xi^a d\xi^b \ecq
  C_3 = -B_2^{\tot} \wedge C_1
  \label{finalC1C3}
\end{align}
in the fundamental domain.
The field strengths $H_3$ and $\tilde{F}_2$ receive no contribution from the discrete connection, but $\tilde{F}_4$ can receive a contribution. Indeed,
\begin{align}
  \tilde{F}_4^{\tot} - \tilde{F}_4^{\bg} &= 
  dC_3 
  - C_1 \wedge H_3^{\bg} 
  - m_0 B_2^{\bg} \wedge B_2
  - \frac{m_0}{2} B_2 \wedge B_2
  \notag \\
  &= - (B_2^{\bg} + B_2) \wedge dC_1
  - m_0 B_2^{\bg} \wedge B_2
  - \frac{m_0}{2} B_2 \wedge B_2
  \notag \\
  &= \frac{m_0}{2} B_2 \wedge B_2
  \ed
  \label{F4mod}
\end{align}
The possible caveat regarding higher-order gauge transformations, mentioned above, applies here as well.
Such transformations can lead to additional obstructions.

\subsection{$B_2$ Example}

We conclude this discussion with two examples of flat $B_2$ connections, taking the manifold to be a twisted 4-torus with monodromy
\begin{align}
  M = \begin{pmatrix}
    1 & N & 0 & 0 \\
    0 & 1 & 0 & 0 \\
    0 & 0 & 1 & N \\
    0 & 0 & 0 & 1
  \end{pmatrix} \ed
\end{align}
Consider the form
\begin{align}
  B_2 = (2\pi)^2 r d\xi^1 \wedge d\xi^3 \ed
\end{align}
We see that
\begin{align}
  f^* B_2 - B_2 = (2\pi)^2 N r \left( 
  d\xi^1 \wedge d\xi^4 + d\xi^2 \wedge d\xi^3
  + N d\xi^2 \wedge d\xi^4
  \right) \ec
\end{align}
Notice that $d\lambda_1 \wedge f^* C_1$ is not closed. Indeed,
\begin{align}
  d\lambda_1 \wedge f^* d C_1 = 
  -m_0 (f^* B_2 - B_2) \wedge f^* B_2
  = (2\pi)^4 m_0 N^2 r^2 d\xi^1 \wedge d\xi^2 \wedge d\xi^3 \wedge d\xi^4 \ed
\end{align}
Therefore, according to the discussion around \eqref{dlfC1}, this connection cannot be turned on when $m_0 \ne 0$.

Next, consider
\begin{align}
  B_2 = (2\pi)^2 r (d\xi^1 \wedge d\xi^4 + d\xi^2 \wedge d\xi^3) \ed
\end{align}
Notice that $B_2 \wedge B_2$ is globally defined, so we expect this connection to contribute to $\tilde{F}_4$.
The quantization conditions follow from
\begin{align}
  f^* (d\xi^1 \wedge d\xi^4) - d\xi^1 \wedge d\xi^4
  &= 
  f^* (d\xi^2 \wedge d\xi^3) - d\xi^2 \wedge d\xi^3
  = N d\xi^2 \wedge d\xi^4 \ed
\end{align}
Notice that both terms in $B_2$ make the same contribution to $f^*B_2 - B_2$, but we must quantize each separately because each term corresponds to a different $\bZ_N$ torsion cycle.
Therefore, the conditions \eqref{btildecond} and \eqref{fbobst} correspond to
\begin{align}
  N r \in \bZ \ecq
  f_0 r \in \bZ \ed
\end{align}
To satisfy the gauge transformation \eqref{fsC13}, we set
\begin{align}
  C_1 = (2\pi)^2 m_0 r (\xi^3 d\xi^2 - \xi^1 d\xi^4) \ed
\end{align}
It is easy to see that $(f^*B_2 - B_2) \wedge f^* C_1 = 0$, and therefore setting $C_3$ as in \eqref{finalC1C3} gives a consistent connection.
Finally, using \eqref{F4mod} we see that the contribution to the 4-form flux is
\begin{align}
  \sqrt{2} \int (\tilde{F}_4^{\tot} - \tilde{F}_4^{\bg}) &= 
  (2\pi)^3 r^2 f_0 \in \frac{(2\pi)^3}{N} \bZ \ed
\end{align}
Using discrete Wilson lines, we find that the flux can be shifted by a non-integer amount, and therefore take on values beyond those allowed by the quantization condition \eqref{FluxQuant}.
The range of accepted values is the same as what we found in section \eqref{C1expl}.

\subsection{Solving the Quantization Conditions}

In this section we solve the quantization conditions for discrete Wilson lines on twisted tori, and write down the general form for such Wilson lines.
This will allow us to compute the general shift in $\tilde{F}_4$ flux due to discrete Wilson lines on such manifolds.

Consider first the condition \eqref{C1DWLquant} for discrete $C_1$ Wilson lines.
Let us write it as a system of linear equations
\begin{align}
  A_{ab} c_b = \sqrt{2} \pi q_a \ecq
  A = M^T - 1 \ecq
  q_a \in \bZ \ed
\end{align}
We would like to find all solutions of this system for arbitrary $q$.
Note that we are not interested in the solutions for a given $q$, but only in the total set of solutions for all possible choices of $q$.
We can therefore redefine $q$ without loss of generality, as long as the redefinition is invertible.
To solve this system we can perform the following row operations on the matrix $A$.
\begin{enumerate}
  \item Add a row times an integer to a different row.
    This corresponds to adding an equation times an integer to another equation, accompanied by a redefinition of $q$.
  \item Multiply a row by -1.
  \item Exchange two rows.
\end{enumerate}
We can also perform the following column operations.
\begin{enumerate}
  \item Add a column times an integer to a different column.
    This corresponds to a redefinition $c_a = c'_a + k c_b$ where $k\in\bZ$ and $a \ne b$.
  \item Multiply a column by -1, corresponding to a redefinition $c_a = -c'_a$.
  \item Exchange two columns, corresponding to a redefinition $c_a \leftrightarrow c_b$.
\end{enumerate}
Using this set of operations, one can reduce $A$ to a diagonal form as follows.
Begin by applying row and column operations such that $A_{11}$ takes the minimal possible positive value.
It is easy to see that $A_{11}$ then divides all elements in the first row and first column, and we can therefore zero out the first row and column except for $A_{11}$.
The claim follows by induction.

Once $A$ is brought to diagonal form, we find the simplified set of constraints
\begin{align}
  N_a c'_a \in \sqrt{2} \pi \bZ \ecq
  c'_a = c_a + \sum_{b \ne a} n_{ab} c_b \ec
\end{align}
where $n_{ab} \in \bZ$.
The integer coefficients $N_a$ correspond to the ranks of the torsion cycles.
A single discrete $C_1$ Wilson line therefore takes the form
\begin{align}
  C_1 = \sqrt{2}\pi \frac{q}{N} 
  \left( d\xi^a + \sum_{b \ne a} k_{ab} d\xi^b \right) \ecq
  k_{ab} \in\bZ \ec
\end{align}
where $q\in\bZ$, and $N$ is the rank of the corresponding torsion cycle.

Let us turn on a flux $H_3 = (2\pi)^2 h \omega_3$, where $h\in\bZ$ and $\omega_3$ is an element in the cohomology. Let us assume that $C_1 \wedge H_3$ is globally defined and non-trivial, so it contributes to the $\tilde{F}_4$ flux.
Within the 4-cycle $\sigma$ that corresponds to $C_1 \wedge H_3$, $H_3$ is a non-trivial 3-form, and the pullback of $C_1$ into $\sigma$ should correspond to its Poincar\'e-dual 1-cycle.
This 1-cycle cannot be a torsion cycle within $\sigma$, because torsion cycles have zero intersection numbers.
Therefore, the pullback of $C_1$ into $\sigma$ should be globally defined within $\sigma$.
But $C_1$ is not invariant under $z\to z-1$ and therefore $\sigma$ cannot wrap $dz$, and $H_3$ cannot include a $dz$ factor.

Let us localize $\sigma$ at $z=0$. 
The pullback of $\omega_3$ into $\sigma$ is then a sum of forms $d\xi^{a'} d\xi^{b'} d\xi^{c'}$ with integer coefficients.
We now see that the discrete Wilson line contribution to the flux can take values in
\begin{align}
  \sqrt{2} \int C_1 \wedge H_3 
  \in \frac{(2\pi)^3}{N} \bZ \ed
\end{align}

Next, let us consider discrete $B_2$ Wilson surfaces. Applying the arguments above to the quantization condition \eqref{btildecond}, we find that $B_2$ takes the form
\begin{align}
  B_2 = (2\pi)^2 \frac{r}{N} (d\xi^i \wedge d\xi^j + \cdots) \ec
\end{align}
where $r\in\bZ$, $N$ is the rank of the discrete Wilson surface, and the extra terms are a sum of forms $d\xi^{i'} \wedge d\xi^{j'}$ (different than $d\xi^i \wedge d\xi^j$) with integer coefficients.
For such a discrete Wilson surface, the condition \eqref{fbobst} implies that
\begin{align}
  \frac{f_0 r}{N} \in \bZ 
  \label{BQfin} \ed
\end{align}
Now, consider a sum of two discrete Wilson surfaces,
\begin{align}
  B_2 = (2\pi)^2 \frac{r}{N} (d\xi^i \wedge d\xi^j + \cdots)
  + (2\pi)^2 \frac{s}{M} (d\xi^k \wedge d\xi^l + \cdots) \ec
\end{align}
and assume that $B_2 \wedge B_2$ is globally defined.
In the previous section we saw that this contributes a piece to the $\tilde{F}_4$ flux, given by \eqref{F4mod}.
Integrating this contribution over any cycle, we find
\begin{align}
  \frac{m_0}{2} \int B_2 \wedge B_2
  = \frac{(2\pi)^3}{\sqrt{2}} \frac{f_0 r s}{N M}
  \int \left( d\xi^i \wedge d\xi^j \wedge d\xi^k \wedge d\xi^l + \cdots \right)
  \ec
\end{align}
where the extra terms are a sum of forms $d\xi^{i'} d\xi^{j'}d\xi^{k'}d\xi^{l'}$ with integer coefficients.
The integral therefore evaluates to an integer.
As for the coefficient in front, applying \eqref{BQfin} to both $r$ and $s$, we see that $NM$ divides both $N f_0 r s$ and $M f_0 r s$. Therefore $NM$ divides $\gcd(N f_0 r s,M f_0 r s) = \gcd(N,M) f_0 r s$.
We find that the discrete $B_2$ Wilson surface contribution to $\tilde{F}_4$ can take values in
\begin{align}
  \frac{m_0}{2} \int B_2 \wedge B_2 \in
  \frac{(2\pi)^3}{\sqrt{2}}
  \frac{1}{\gcd(N,M)} \bZ \ed
\end{align}

\subsection{Combinations of Fluxes and Flat Connections}

In the previous sections we considered concrete examples of the general arguments made in section \ref{gc}.
We have not covered all the possible combinations of fluxes and discrete Wilson lines.
Though it is beyond the scope of the present work to consider all such combinations, the considerations of section \ref{gc} lead us to expect that such combinations can only lead to additional obstructions, rather than to new qualitative features beyond those we already found.
We expect to encounter obstructions because backgrounds that are not globally defined modify the gauge transformations. For example, a choice such as $C_3 = -B_2^\bg \wedge C_1$ \eqref{C3forC1} will no longer be sufficient to account for the transformation \eqref{C1lineC3t} when $B_2^\bg$ is not globally defined, and it may not be possible to evade this problem by turning on additional gauge transformations $C_3 \to C_3 + d\Lambda_2$.


\begin{thebibliography}{99}

\bibitem{Silverstein:2008sg} 
  E.~Silverstein and A.~Westphal,
  ``Monodromy in the CMB: Gravity Waves and String Inflation,''
  Phys.\ Rev.\ D {\bf 78}, 106003 (2008)
  [\href{http://arxiv.org/abs/0803.3085}{arXiv:0803.3085} [hep-th]].

\bibitem{Baumann:2009ds} 
  D.~Baumann,
  ``TASI Lectures on Inflation,''
  [\href{http://arxiv.org/abs/0907.5424}{arXiv:0907.5424} [hep-th]].

\bibitem{Lyth:1996im} 
  D.~H.~Lyth,
  ``What would we learn by detecting a gravitational wave signal in the cosmic microwave background anisotropy?,''
  Phys.\ Rev.\ Lett.\  {\bf 78}, 1861 (1997)
  [\href{http://arxiv.org/abs/hep-ph/9606387}{hep-ph/9606387}].

\bibitem{Hotchkiss:2011gz} 
  S.~Hotchkiss, A.~Mazumdar and S.~Nadathur,
  ``Observable gravitational waves from inflation with small field excursions,''
  JCAP {\bf 1202}, 008 (2012)
  [\href{http://arxiv.org/abs/arXiv:1110.5389}{arXiv:1110.5389} [astro-ph.CO]].

\bibitem{Kachru:2003sx} 
  S.~Kachru, R.~Kallosh, A.~D.~Linde, J.~M.~Maldacena, L.~P.~McAllister and S.~P.~Trivedi,
  ``Towards inflation in string theory,''
  JCAP {\bf 0310}, 013 (2003)
  [\href{http://arxiv.org/abs/hep-th/0308055}{hep-th/0308055}].

\bibitem{Burgess:2001fx} 
  C.~P.~Burgess, M.~Majumdar, D.~Nolte, F.~Quevedo, G.~Rajesh and R.~-J.~Zhang,
  ``The Inflationary brane anti-brane universe,''
  JHEP {\bf 0107}, 047 (2001)
  [\href{http://arxiv.org/abs/hep-th/0105204}{hep-th/0105204}].

\bibitem{Dvali:2001fw} 
  G.~R.~Dvali, Q.~Shafi and S.~Solganik,
  ``D-brane inflation,''
  [\href{http://arxiv.org/abs/hep-th/0105203}{hep-th/0105203}].

\bibitem{McAllister:2008hb} 
  L.~McAllister, E.~Silverstein and A.~Westphal,
  ``Gravity Waves and Linear Inflation from Axion Monodromy,''
  Phys.\ Rev.\ D {\bf 82}, 046003 (2010)
  [\href{http://arxiv.org/abs/0808.0706}{arXiv:0808.0706} [hep-th]].

\bibitem{Silverstein:2007ac} 
  E.~Silverstein,
  ``Simple de Sitter Solutions,''
  Phys.\ Rev.\ D {\bf 77}, 106006 (2008)
  [\href{http://arxiv.org/abs/0712.1196}{arXiv:0712.1196} [hep-th]].

\bibitem{Hertzberg:2007wc} 
  M.~P.~Hertzberg, S.~Kachru, W.~Taylor and M.~Tegmark,
  ``Inflationary Constraints on Type IIA String Theory,''
  JHEP {\bf 0712}, 095 (2007)
  [\href{http://arxiv.org/abs/0711.2512}{arXiv:0711.2512} [hep-th]].

\bibitem{Caviezel:2008ik} 
  C.~Caviezel, P.~Koerber, S.~Kors, D.~Lust, D.~Tsimpis and M.~Zagermann,
  ``The Effective theory of type IIA AdS(4) compactifications on nilmanifolds and cosets,''
  Class.\ Quant.\ Grav.\  {\bf 26}, 025014 (2009)
  [\href{http://arxiv.org/abs/0806.3458}{arXiv:0806.3458} [hep-th]].

\bibitem{Caviezel:2008tf} 
  C.~Caviezel, P.~Koerber, S.~Kors, D.~Lust, T.~Wrase and M.~Zagermann,
  ``On the Cosmology of Type IIA Compactifications on SU(3)-structure Manifolds,''
  JHEP {\bf 0904}, 010 (2009)
  [\href{http://arxiv.org/abs/0812.3551}{arXiv:0812.3551} [hep-th]].

\bibitem{Flauger:2008ad} 
  R.~Flauger, S.~Paban, D.~Robbins and T.~Wrase,
  ``Searching for slow-roll moduli inflation in massive type IIA supergravity with metric fluxes,''
  Phys.\ Rev.\ D {\bf 79}, 086011 (2009)
  [\href{http://arxiv.org/abs/0812.3886}{arXiv:0812.3886} [hep-th]].

\bibitem{Danielsson:2009ff} 
  U.~H.~Danielsson, S.~S.~Haque, G.~Shiu and T.~Van Riet,
  ``Towards Classical de Sitter Solutions in String Theory,''
  JHEP {\bf 0909}, 114 (2009)
  [\href{http://arxiv.org/abs/0907.2041}{arXiv:0907.2041} [hep-th]].

\bibitem{deCarlos:2009fq} 
  B.~de Carlos, A.~Guarino and J.~M.~Moreno,
  ``Flux moduli stabilisation, Supergravity algebras and no-go theorems,''
  JHEP {\bf 1001}, 012 (2010)
  [\href{http://arxiv.org/abs/0907.5580}{arXiv:0907.5580} [hep-th]].

\bibitem{deCarlos:2009qm} 
  B.~de Carlos, A.~Guarino and J.~M.~Moreno,
  ``Complete classification of Minkowski vacua in generalised flux models,''
  JHEP {\bf 1002}, 076 (2010)
  [\href{http://arxiv.org/abs/0911.2876}{arXiv:0911.2876} [hep-th]].

\bibitem{Wrase:2010ew} 
  T.~Wrase and M.~Zagermann,
  ``On Classical de Sitter Vacua in String Theory,''
  Fortsch.\ Phys.\  {\bf 58}, 906 (2010)
  [\href{http://arxiv.org/abs/1003.0029}{arXiv:1003.0029} [hep-th]].

\bibitem{Danielsson:2010bc} 
  U.~H.~Danielsson, P.~Koerber and T.~Van Riet,
  ``Universal de Sitter solutions at tree-level,''
  JHEP {\bf 1005}, 090 (2010)
  [\href{http://arxiv.org/abs/1003.3590}{arXiv:1003.3590} [hep-th]].

\bibitem{Andriot:2010ju} 
  D.~Andriot, E.~Goi, R.~Minasian and M.~Petrini,
  ``Supersymmetry breaking branes on solvmanifolds and de Sitter vacua in string theory,''
  JHEP {\bf 1105}, 028 (2011)
  [\href{http://arxiv.org/abs/1003.3774}{arXiv:1003.3774} [hep-th]].

\bibitem{Danielsson:2011au} 
  U.~H.~Danielsson, S.~S.~Haque, P.~Koerber, G.~Shiu, T.~Van Riet and T.~Wrase,
  ``De Sitter hunting in a classical landscape,''
  Fortsch.\ Phys.\  {\bf 59}, 897 (2011)
  [\href{http://arxiv.org/abs/1103.4858}{arXiv:1103.4858} [hep-th]].

\bibitem{Freund:1980xh} 
  P.~G.~O.~Freund and M.~A.~Rubin,
  ``Dynamics of Dimensional Reduction,''
  Phys.\ Lett.\ B {\bf 97}, 233 (1980).

\bibitem{Scherk:1979zr} 
  J.~Scherk and J.~H.~Schwarz,
  ``How to Get Masses from Extra Dimensions,''
  Nucl.\ Phys.\ B {\bf 153}, 61 (1979).

\bibitem{Bergshoeff:1996ui} 
  E.~Bergshoeff, M.~de Roo, M.~B.~Green, G.~Papadopoulos and P.~K.~Townsend,
  ``Duality of type II 7 branes and 8 branes,''
  Nucl.\ Phys.\ B {\bf 470}, 113 (1996)
  [\href{http://arxiv.org/abs/hep-th/9601150}{hep-th/9601150}].

\bibitem{Cowdall:1996tw} 
  P.~M.~Cowdall, H.~Lu, C.~N.~Pope, K.~S.~Stelle and P.~K.~Townsend,
  ``Domain walls in massive supergravities,''
  Nucl.\ Phys.\ B {\bf 486}, 49 (1997)
  [\href{http://arxiv.org/abs/hep-th/9608173}{hep-th/9608173}].

\bibitem{Kaloper:1999yr} 
  N.~Kaloper and R.~C.~Myers,
  ``The Odd story of massive supergravity,''
  JHEP {\bf 9905}, 010 (1999)
  [\href{http://arxiv.org/abs/hep-th/9901045}{hep-th/9901045}].

\bibitem{Cvetic:2003jy} 
  M.~Cvetic, G.~W.~Gibbons, H.~Lu and C.~N.~Pope,
  ``Consistent group and coset reductions of the bosonic string,''
  Class.\ Quant.\ Grav.\  {\bf 20}, 5161 (2003)
  [\href{http://arxiv.org/abs/hep-th/0306043}{hep-th/0306043}].

\bibitem{Gurrieri:2002wz} 
  S.~Gurrieri, J.~Louis, A.~Micu and D.~Waldram,
  ``Mirror symmetry in generalized Calabi-Yau compactifications,''
  Nucl.\ Phys.\ B {\bf 654}, 61 (2003)
  [\href{http://arxiv.org/abs/hep-th/0211102}{hep-th/0211102}].

\bibitem{Gurrieri:2002iw} 
  S.~Gurrieri and A.~Micu,
  ``Type IIB theory on half flat manifolds,''
  Class.\ Quant.\ Grav.\  {\bf 20}, 2181 (2003)
  [\href{http://arxiv.org/abs/hep-th/0212278}{hep-th/0212278}].

\bibitem{Schulz:2004ub} 
  M.~B.~Schulz,
  ``Superstring orientifolds with torsion: O5 orientifolds of torus fibrations and their massless spectra,''
  Fortsch.\ Phys.\  {\bf 52}, 963 (2004)
  [\href{http://arxiv.org/abs/hep-th/0406001}{hep-th/0406001}].

\bibitem{Schulz:2004tt} 
  M.~B.~Schulz,
  ``Calabi-Yau duals of torus orientifolds,''
  JHEP {\bf 0605}, 023 (2006)
  [\href{http://arxiv.org/abs/hep-th/0412270}{hep-th/0412270}].

\bibitem{Derendinger:2004jn} 
  J.~-P.~Derendinger, C.~Kounnas, P.~M.~Petropoulos and F.~Zwirner,
  ``Superpotentials in IIA compactifications with general fluxes,''
  Nucl.\ Phys.\ B {\bf 715}, 211 (2005)
  [\href{http://arxiv.org/abs/hep-th/0411276}{hep-th/0411276}].

\bibitem{Derendinger:2005ph} 
  J.~-P.~Derendinger, C.~Kounnas, P.~M.~Petropoulos and F.~Zwirner,
  ``Fluxes and gaugings: N=1 effective superpotentials,''
  Fortsch.\ Phys.\  {\bf 53}, 926 (2005)
  [\href{http://arxiv.org/abs/hep-th/0503229}{hep-th/0503229}].

\bibitem{Dall'Agata:2005ff} 
  G.~Dall'Agata and S.~Ferrara,
  ``Gauged supergravity algebras from twisted tori compactifications with fluxes,''
  Nucl.\ Phys.\ B {\bf 717}, 223 (2005)
  [\href{http://arxiv.org/abs/hep-th/0502066}{hep-th/0502066}].

\bibitem{Villadoro:2005cu} 
  G.~Villadoro and F.~Zwirner,
  ``N=1 effective potential from dual type-IIA D6/O6 orientifolds with general fluxes,''
  JHEP {\bf 0506}, 047 (2005)
  [\href{http://arxiv.org/abs/hep-th/0503169}{hep-th/0503169}].

\bibitem{Camara:2005dc} 
  P.~G.~Camara, A.~Font and L.~E.~Ibanez,
  ``Fluxes, moduli fixing and MSSM-like vacua in a simple IIA orientifold,''
  JHEP {\bf 0509}, 013 (2005)
  [\href{http://arxiv.org/abs/hep-th/0506066}{hep-th/0506066}].

\bibitem{Hull:2005hk} 
  C.~M.~Hull and R.~A.~Reid-Edwards,
  ``Flux compactifications of string theory on twisted tori,''
  Fortsch.\ Phys.\  {\bf 57}, 862 (2009)
  [\href{http://arxiv.org/abs/hep-th/0503114}{hep-th/0503114}].

\bibitem{Fre':2006ut} 
  P.~Fre' and M.~Trigiante,
  ``Twisted tori and fluxes: A No go theorem for Lie groups of weak G(2) holonomy,''
  Nucl.\ Phys.\ B {\bf 751}, 343 (2006)
  [\href{http://arxiv.org/abs/hep-th/0603011}{hep-th/0603011}].

\bibitem{ReidEdwards:2009nu} 
  R.~A.~Reid-Edwards,
  ``Flux compactifications, twisted tori and doubled geometry,''
  JHEP {\bf 0906}, 085 (2009)
  [\href{http://arxiv.org/abs/0904.0380}{arXiv:0904.0380} [hep-th]].

\bibitem{Haque:2008jz} 
  S.~S.~Haque, G.~Shiu, B.~Underwood and T.~Van Riet,
  ``Minimal simple de Sitter solutions,''
  Phys.\ Rev.\ D {\bf 79}, 086005 (2009)
  [\href{http://arxiv.org/abs/0810.5328}{arXiv:0810.5328} [hep-th]].

\bibitem{Douglas:2010rt} 
  M.~R.~Douglas and R.~Kallosh,
  ``Compactification on negatively curved manifolds,''
  JHEP {\bf 1006}, 004 (2010)
  [\href{http://arxiv.org/abs/1001.4008}{arXiv:1001.4008} [hep-th]].

\bibitem{Dong:2010pm} 
  X.~Dong, B.~Horn, E.~Silverstein and G.~Torroba,
  ``Micromanaging de Sitter holography,''
  Class.\ Quant.\ Grav.\  {\bf 27}, 245020 (2010)
  [\href{http://arxiv.org/abs/1005.5403}{arXiv:1005.5403} [hep-th]].

\bibitem{Blaback:2010sj} 
  J.~Blaback, U.~H.~Danielsson, D.~Junghans, T.~Van Riet, T.~Wrase and M.~Zagermann,
  ``Smeared versus localised sources in flux compactifications,''
  JHEP {\bf 1012}, 043 (2010)
  [\href{http://arxiv.org/abs/1009.1877}{arXiv:1009.1877} [hep-th]].

\bibitem{Culver:1966aa}
  W.~J.~Culver,
  ``On the existence and uniqueness of the real logarithm of a matrix'',
  Proc.\ Amer.\ Math.\ Soc.\ {\bf 17} (1966), 1146-1151

\bibitem{Dong:2010in} 
  X.~Dong, B.~Horn, E.~Silverstein and A.~Westphal,
  ``Simple exercises to flatten your potential,''
  Phys.\ Rev.\ D {\bf 84}, 026011 (2011)
  [\href{http://arxiv.org/abs/1011.4521}{arXiv:1011.4521} [hep-th]].

\bibitem{Gurari:2010aa}
  G.~Gur-Ari,
  ``Large-Field Inflation from D-Branes on $T^5$ Bundles'',
  M.Sc. thesis, Weizmann Institute of Science (2010).

\bibitem{Grana:2008yw} 
  M.~Grana, R.~Minasian, M.~Petrini and D.~Waldram,
  ``T-duality, Generalized Geometry and Non-Geometric Backgrounds,''
  JHEP {\bf 0904}, 075 (2009)
  [\href{http://arxiv.org/abs/0807.4527}{arXiv:0807.4527} [hep-th]].

\bibitem{Romans:1985tz} 
  L.~J.~Romans,
  ``Massive N=2a Supergravity in Ten-Dimensions,''
  Phys.\ Lett.\ B {\bf 169}, 374 (1986).

\bibitem{DeWolfe:2005uu} 
  O.~DeWolfe, A.~Giryavets, S.~Kachru and W.~Taylor,
  ``Type IIA moduli stabilization,''
  JHEP {\bf 0507}, 066 (2005)
  [\href{http://arxiv.org/abs/hep-th/0505160}{hep-th/0505160}].

\bibitem{Green:1996bh} 
  M.~B.~Green, C.~M.~Hull and P.~K.~Townsend,
  ``D-brane Wess-Zumino actions, t duality and the cosmological constant,''
  Phys.\ Lett.\ B {\bf 382}, 65 (1996)
  [\href{http://arxiv.org/abs/hep-th/9604119}{hep-th/9604119}].

\bibitem{Bergshoeff:2003ri} 
  E.~Bergshoeff, U.~Gran, R.~Linares, M.~Nielsen, T.~Ortin and D.~Roest,
  ``The Bianchi classification of maximal D = 8 gauged supergravities,''
  Class.\ Quant.\ Grav.\  {\bf 20}, 3997 (2003)
  [\href{http://arxiv.org/abs/hep-th/0306179}{hep-th/0306179}].

\bibitem{Gross:1983hb} 
  D.~J.~Gross and M.~J.~Perry,
  ``Magnetic Monopoles in Kaluza-Klein Theories,''
  Nucl.\ Phys.\ B {\bf 226}, 29 (1983).

\bibitem{Giveon:2008zn} 
  A.~Giveon and D.~Kutasov,
  ``Seiberg Duality in Chern-Simons Theory,''
  Nucl.\ Phys.\ B {\bf 812}, 1 (2009)
  [\href{http://arxiv.org/abs/0808.0360}{arXiv:0808.0360} [hep-th]].

\bibitem{Frey:2002hf} 
  A.~R.~Frey and J.~Polchinski,
  ``N=3 warped compactifications,''
  Phys.\ Rev.\ D {\bf 65}, 126009 (2002)
  [\href{http://arxiv.org/abs/hep-th/0201029}{hep-th/0201029}].

\bibitem{polchinski1998string}
  J.~Polchinski,
  String Theory, Vol. II (1998)

\bibitem{Buscher:1987qj} 
  T.~H.~Buscher,
  ``Path Integral Derivation of Quantum Duality in Nonlinear Sigma Models,''
  Phys.\ Lett.\ B {\bf 201}, 466 (1988).

\bibitem{Kachru:2002sk} 
  S.~Kachru, M.~B.~Schulz, P.~K.~Tripathy and S.~P.~Trivedi,
  ``New supersymmetric string compactifications,''
  JHEP {\bf 0303}, 061 (2003)
  [\href{http://arxiv.org/abs/hep-th/0211182}{hep-th/0211182}].

\bibitem{unpub} 
  G.~Gur-Ari, to appear.

\bibitem{Nakahara} 
  M.~Nakahara, 
  ``Geometry, Topology, and Physics,''
  Institute of Physics Publishing (2003)

\bibitem{Townsend:1995gp} 
  P.~K.~Townsend,
  ``P-brane democracy,''
  In *Duff, M.J. (ed.): The world in eleven dimensions* 375-389
  [\href{http://arxiv.org/abs/hep-th/9507048}{hep-th/9507048}].

\bibitem{Bergshoeff:2001pv} 
  E.~Bergshoeff, R.~Kallosh, T.~Ortin, D.~Roest and A.~Van Proeyen,
  ``New formulations of D = 10 supersymmetry and D8 - O8 domain walls,''
  Class.\ Quant.\ Grav.\  {\bf 18}, 3359 (2001)
  [\href{http://arxiv.org/abs/hep-th/0103233}{hep-th/0103233}].

\bibitem{Douglas:1995bn} 
  M.~R.~Douglas,
  ``Branes within branes,''
  In *Cargese 1997, Strings, branes and dualities* 267-275
  [\href{http://arxiv.org/abs/hep-th/9512077}{hep-th/9512077}].

\end{thebibliography}
\end{document}